\documentclass[11pt]{article}
\pdfoutput=1
\usepackage{jheppubme}
\usepackage[table]{xcolor}
\usepackage{graphicx, psfrag}
\usepackage{float}
\usepackage{caption}
\usepackage[mathscr]{euscript}
\usepackage{dynkin-diagrams}
\usepackage[toc,page]{appendix}
\usepackage{hyperref}
\usepackage{color}
\usepackage{booktabs,caption}
\usepackage{lscape}
\definecolor{Mygrey}{gray}{0.8}
\definecolor{Mywhite}{gray}{1.0}

\newcommand{\be}{\begin{equation}}
\newcommand{\ee}{\end{equation}}
\newcommand{\bea}{\begin{eqnarray}}
\newcommand{\eea}{\end{eqnarray}}
\usepackage{amsmath,amssymb,epsfig,jheppubme}
\usepackage{multirow,longtable,enumerate,bm}
\usepackage[flushleft]{threeparttable}

\newcommand\half{\frac12}

\newcommand\bi{\begin{itemize}}
\newcommand\ei{\end{itemize}}

\newcommand\tl{{\tilde \ell}}

\newcommand\tc{{\tilde c}}

\newcommand\tilh{{\tilde h}}
\newcommand\tchi{{\tilde \chi}}

\newcommand\hchi{{\hat \chi}}

\newcommand\btau{{\bar \tau}}

\newcommand\cN{{\cal N}}

\newcommand\cC{{\cal C}}
\newcommand\tcC{{\tilde{\cal C}}}

\newcommand{\cH}{{\cal H}}

\newcommand\bchi{{\overline \chi}}

\newcommand\ZZ{\hbox{Z\kern-.4emZ}}
\newcommand\sZZ{\hbox{\sevenfont Z\kern-.4emZ}}

\newcommand{\eref}[1]{Eq.\,(\ref{#1})}
\newcommand{\Comment}[1]{{}}

\definecolor{Mygrey}{gray}{0.8}
\definecolor{Mywhite}{gray}{1.0}

\colorlet{dred}{red!70!black!100!}

\def\IB{\relax{\rm I\kern-.18em B}}

\def\ID{\relax{\rm I\kern-.18em D}}
\def\IE{\relax{\rm I\kern-.18em E}}
\def\IF{\relax{\rm I\kern-.18em F}}
\def\II{\relax{\rm I\kern-.18em I}}

\def\Id{\relax{1\kern-.32em 1}}
\def\IG{\relax\hbox{$\inbar\kern-.3em{\rm G}$}}
\def\IR{\relax{\rm I\kern-.18em R}}

\title{Results of (3,0) GHM Duals}

\title{New meromorphic CFTs from cosets} 
\author[a]{Arpit Das,}
\author[b,c]{Chethan N. Gowdigere} 
\author[d]{and Sunil Mukhi}

\affiliation[a]{Centre for Particle Theory, Department of Mathematical Sciences,\\
Durham University, South Road, Durham DH1 3LE, United Kingdom}
\affiliation[b]{National Institute of Science Education and Research Bhubaneshwar,\\
P.O. Jatni, Khurda, 752050, Odisha, India}
\affiliation[c]{Homi Bhabha National Institute, Training School Complex,\\
Anushakti Nagar, Mumbai 400094, India}
\affiliation[d]{Indian Institute of Science Education and Research,\\ Homi Bhabha Rd, Pashan, Pune 411 008, India}

\emailAdd{arpit.das@durham.ac.uk}
\emailAdd{chethan.gowdigere@niser.ac.in}
\emailAdd{sunil.mukhi@gmail.com}
    
\abstract{In recent years it has been understood that new rational CFTs can be discovered by applying the coset construction to meromorphic CFTs. Here we turn this approach around and show that the coset construction, together with the classification of meromorphic CFT with $c\le 24$, can be used to predict the existence of new meromorphic CFTs with $c\ge 32$ whose Kac-Moody algebras are non-simply-laced and/or at levels greater than 1. This implies they are non-lattice theories. Using three-character coset relations, we propose 34 infinite series of meromorphic theories with arbitrarily large central charge, as well as 46 theories at $c=32$ and $c=40$.}

\preprint{}


\begin{document}

\maketitle

\section{Introduction}
\label{Introduction}

A rational 2d conformal field theory (RCFT) is characterised by having a finite set of holomorphic characters $\chi_i(\tau)$ and a partition function of the form:
\be
\mathcal{Z}(\tau,\btau)=\sum_{i,j=0}^{n-1}M_{ij}\,\bchi_i(\btau)\chi_j(\tau) \label{zij}
\ee
where $n$ is the total number of linearly independent characters of some chiral algebra.

A classification programme for rational conformal field theories in 2d was initiated in \cite{Mathur:1988rx,Mathur:1988na}. It is based on the fact that characters are vector-valued forms that solve a modular linear differential equation (MLDE) whose order is the number of characters. Such equations have finitely many parameters and these can be varied to scan for solutions that satisfy the basic criteria to be those of a conformal field theory. We will refer to such solutions as ``admissible characters''. From the study of MLDE it emerged that an important classifier for RCFT with a given number of characters is an integer $\ell\ge 0,\ell\ne 1$ called the Wronskian index (for a detailed review, see \cite{Das:2020wsi}). Admissible characters for bosonic CFTs have been constructed in \cite{Naculich:1988xv,Mathur:1988gt,Hampapura:2015cea,Harvey:2018rdc,Chandra:2018pjq,Mukhi:2020gnj,Das:2021uvd,Kaidi:2021ent,Duan:2022ltz} and for fermionic CFTs in \cite{Durganandini:1989es,Bae:2020xzl,Bae:2021mej}.

Admissible characters do not, in general, correspond to any RCFT. For meromorphic theories (those with a single character, $n=1$) this is particularly obvious. At $c=24$ there is an infinite set of admissible characters as we will see below, but only a finite subset corresponds to a CFT as shown in the seminal work of Schellekens \cite{Schellekens:1992db}. Moreover a given character in this finite set sometimes corresponds to multiple different CFTs. That an analogous phenomenon holds for theories with multiple characters was made explicit in recent times \cite{Harvey:2018rdc,Chandra:2018pjq,Mukhi:2020gnj} with the discovery of infinite families of admissible characters depending on unbounded integers.

The classification of RCFTs for a fixed number of characters and Wronskian index thus requires two steps: the classification of admissible characters, and the sub-classification of those that describe genuine RCFT. Both steps are easiest for small values of $n$ as well as the Wronskian index $\ell$, because in this case the characters are completely determined by their critical exponents \cite{Mathur:1988na,Mathur:1988gt}. Given a set of admissible characters, a useful method to find corresponding CFT comes from the coset construction of \cite{Goddard:1984vk,Goddard:1986ee} as implemented for meromorphic numerator theories in \cite{Gaberdiel:2016zke}\nobreak \footnote{See \cite{Schellekens:1988ag,Moore:1989vd} for earlier discussions of meromorphic cosets.}. This works as follows. We pick a known meromorphic CFT $\cal H$ as well as a known RCFT ${\cal C}$ with $n$ characters and suitable exponents, such that the coset: 
\be
{\tilde{\cal C}}=\frac{\cal H}{\cal C}
\label{cosrel}
\ee
obtained by embedding the Kac-Moody algebra of $\cC$ in that of $\cH$  corresponds to the given admissible characters. Whenever this can be done, we may conclude that these characters describe the CFT $\tcC$. 
This approach has been implemented in \cite{Gaberdiel:2016zke,Chandra:2018ezv,MR:upcoming,DGM:upcoming} where $\cH$ is a known CFT at $c=8,16,24$ (these are completely classified) or a lattice theory at $c=32$ (these are too many to classify but many lattice theories are known \cite{kervaire1994unimodular,king2003mass}). In particular, in \cite{Gaberdiel:2016zke} this procedure was first used to construct the CFTs' corresponding to admissible characters with $(n,\ell)=(2,2)$ that had originally been found nearly three decades earlier \cite{Naculich:1988xv} but had not been previously identified with CFTs. 

Let us briefly review some basic aspects of the meromorphic coset relation (more details can be found in \cite{Gaberdiel:2016zke,Das:2021uvd}). The numerator theory $\cH$ is typically an extension of a non-simple Kac-Moody algebra $\oplus_i {\cal G}_{r_i,k_i}$ by higher-spin generators that organise a subset of Kac-Moody characters into a single character. We denote these theories by ${\cal E}_1[\oplus_i {\cal G}_{r_i,k_i}]$ where $\cal E$ stands for ``extension'' and the subscript 1 tells us that the extended theory has a single character. There are broadly two types of meromorphic theories: those corresponding to free bosons on an even, self-dual lattice, for which the Kac-Moody algebra only contains simply-laced factors ($A_r,D_r,E_6,E_7,E_8$) at level 1, and the rest, which we call non-lattice theories. These are characterised by the presence of  non-simply-laced factors ($B_r,C_r,F_4,G_2$) and/or levels greater than 1. Some non-lattice theories can be derived as orbifolds of lattice theories, while others are more complicated to construct \cite{goddard1989meromorphic,Schellekens:1992db,dolan1996conformal}. The denominator theories $\cC$, at least in the original references \cite{Gaberdiel:2016zke,Hampapura:2016mmz,Chandra:2018pjq}, are taken to be affine theories belonging to a current-algebra minimal series  \cite{Knizhnik:1984nr} or occasionally the Virasoro minimal series ${\cal M}(p,q)$ \cite{Belavin:1984vu}. The coset theories $\tcC$ again typically have non-simple Kac-Moody algebras that are extended by other chiral generators so that they have a smaller number $n$ of characters than the affine theories for the same algebras. Following the notation introduced above for meromorphic theories, we denote these by ${\cal E}_n[\oplus_i {\cal G}_{r_i,k_i}]$ where the subscript $n$ denotes the number of characters.

If the characters of ${\cal C}$ are denoted $\chi_i(\tau)$, those of $\tcC$  are denoted $\tchi_i$ and the single character of $\cal H$ is denoted $\chi^{\cal H}$, then the coset relation is embodied in a holomorphic bilinear relation:
\be
\sum_{i=0}^{n-1}d_i\,\chi_i(\tau)\,\tchi_i(\tau)=\chi^{\cal H}(\tau)
\label{bilinrel}
\ee
where $d_i$ are positive integers and $d_0=1$.

When both $\cH$ and $\cal C$ correspond to CFTs having a Sugawara stress tensor in terms of Kac-Moody currents, then by embedding the currents of $\cal C$ in those of $\cH$ one defines the stress tensor of the coset theory $\tilde {\cal C}$. At a physics level of rigour, we take this to be a proof that $\cC$ is a genuine CFT \cite{Gaberdiel:2016zke}. But it is also possible for \eref{bilinrel} to be satisfied when one or both of $\cC,\tcC$ does not 
have any Kac-Moody currents. Also there are cases where none of $\cC,\tcC,\cH$ has Kac-Moody currents (for example, $\cH$ may correspond to the Monster module \cite{Frenkel:1988xz,Borcherds:1}). In such cases the coset construction of \cite{Gaberdiel:2016zke} is more tricky to apply, since without a Sugawara construction we do not have an explicit expression for the stress tensor of $\cC$. At the same time, the bilinear relation can certainly be verified as easily as for the Sugawara case. 
So it is compelling to believe that, even in the absence of Sugawara stress tensors, if the bilinear relation holds and $\cH$ and $\cC$ are both CFTs, then so is $\tcC$. One such example \cite{Hampapura:2016mmz} arises when $\cC$ is the Ising model and $\tcC$ is the Baby Monster CFT \cite{Hoehn:Baby8}.

As described above, the coset relation \eref{cosrel} has been used to find interesting theories $\tcC$ given meromorphic theories $\cH$ and known denominator CFTs $\cC$ to divide them by. As input, this method has relied on known classifications of meromorphic theories at $c=8,16,24$ \cite{goddard1989meromorphic,Schellekens:1992db} as well as special classes of lattices at $c=32$ \cite{kervaire1994unimodular,king2003mass} to find new $n$-character CFTs. However, if $\cC$ and $\tcC$ are both known to describe CFTs then this relation can be used in the other direction: to argue that $\cH$ also corresponds to a CFT. Thus, in principle new meromorphic theories can be found in this way.

In the present work we propose a practical way to carry this out. The key idea is to first generate an extension-type CFT $\tcC$ using the coset construction, with $\cH$ being a known meromorphic theory in the Schellekens list \cite{Schellekens:1992db} and $\cC$ being a suitable affine theory embedded in it. Next, we remove $\cC$ from the story and find bilinear relations where $\tcC$ is paired with the characters of another known CFT $\cC'$. For $n=3$, large numbers of such coset pairs were generated in \cite{Das:2021uvd}. Since $\cC'$ is a different theory from $\cC$ (and not a tensor product of $\cC$ with something else), the result is a {\em new} meromorphic theory at a different central charge.

The affine algebra of the new meromorphic theory is the sum of algebras of $\cC'$ and $\tcC$, denoted $\mathcal{E}_1[\mathcal{C}'\oplus\Tilde{\mathcal{C}}]$, of the multi-character theory $\mathcal{C}'\oplus\Tilde{\mathcal{C}}$. This extension arises from the bilinear relation satisfied by $\cC'$ and $\tcC$. In each case we explicitly find the character of $\cH$. In most cases it will turn out that $\cC'\oplus \tcC$ is non-simply-laced and/or has factors of level $>1$, showing that it must be a non-lattice theory.

Using the above approach we predict the existence of 34 infinite families of novel meromorphic theories with central charges $c^{\mathcal{H}}=8(m+3)$ where $m$ is an integer $\ge 1$. Every such infinite family starts with a member at $c^{\mathcal{H}}=24$ that can be found in Schellekens' list \cite{Schellekens:1992db}. We also predict the existence of 46 novel meromorphic theories with central charges $c^{\mathcal{H}}=32$ and $c^{\mathcal{H}}=40$.  We now turn to a description of the method and then present our results.

\section{Background}

We start by briefly summarising the admissible characters for meromorphic theories at $c^{\mathcal{H}}=8N$ where $N$ is an integer $\geq 1$. From the theory of modular forms (see for example \cite{murty2016problems}) we know that any polynomial \footnote{Actually any rational function of $j(q)$ would be invariant, but we are specialising to polynomials as we want these functions to be holomorphic in $\mathbb{H} \setminus \{i\infty\}$.} of the Klein $j$-invariant $j(q)$ is a modular invariant function. The $j$-invariant has the expansion:
\be
j(q)=q^{-1}+744+196884q+\cdots
\ee
If we are willing to accept modular invariance up to a phase then the polynomial can be multiplied by $j(q)^{\frac{1}{3}}$ or $j(q)^{\frac{2}{3}}$. Thus the most general modular invariant is of the form:
\be
    \chi^\cH(\tau) = j^{\frac{N}{3}-s}(j^s + a_1\,j^{s-1}+ a_2 \, j^{s-2} + \ldots\ldots + a_s),
\label{gen_j}
\ee
where $s := \left \lfloor \frac{N}{3} \right \rfloor$.

For $c^\cH=24,32,40$, the corresponding characters are of the form:
\begin{align}
    &c^{\mathcal{H}}=24: \, \, \, \chi^\cH(\tau) = j+\cN, \label{8mero} \\
    &c^{\mathcal{H}}=32: \, \, \, \chi^\cH(\tau) = j^{\frac{1}{3}}(j+\cN) \label{16mero} \\
    &c^{\mathcal{H}}=40: \, \, \, \chi^\cH(\tau) = j^{\frac{2}{3}}(j + \cN), \label{24mero0}
\end{align}
where we have renamed $a_1$ as $\cN$. In the $c^\cH=24$ case we see that all values of $\cN\ge -744$ are admissible, but it is known that only a finite subset correspond to CFTs. \cite{Schellekens:1992db}.

For $n$-character RCFT, the characters are vector-valued modular forms that have the general form:
\be
\chi_i(q)=q^{-\frac{c}{24}}(a_{i,0}+a_{i,1}q+a_{i,2}q^2+\cdots),~~i=0,1,\cdots,n-1
\ee
For the characters to be admissible, the coefficients $a_{i,r}$ must be non-negative integers. Moreover we must have $a_{0,0}=1$ (non-degeneracy of the identity character). We also define $D_i=a_{i,0}, m_1=a_{0,1}, m_2=a_{0,2}$. These are, respectively, the ground-state degeneracy of each of the generalised primaries, the number of spin-1 currents and the number of spin-2 currents. 

For $n=2$ there is an explicit and complete classification of all admissible characters \cite{Chandra:2018pjq}. However for $n\ge 3$ the complete classification remains an open problem. Nevertheless, admissible characters can be found by writing a general Modular Linear Differential Equation of the form:
\be
\big(D^n+\phi_2(\tau) D^{n-1}+\cdots +\phi_{2n}(\tau)\big)\chi=0
\ee
where $D$ is a covariant derivative on torus moduli space and $\phi_{2j}(\tau)$ are meromorphic modular functions of weight $2j$. The maximum number of poles of the $\phi_{2r}$ is called the Wronskian index $\ell$. Consider the above equation for a fixed order $n$ and a fixed value of $\ell$. In this case the set of $\phi_{2j}$ depends on a finite number of parameters, and one scans the parameter space to find those values where the coefficients $a_{i,r}$ satisfy the admissibility criteria. Of relevance in the present work will be solutions for $n=3$ and $\ell=0$, in which case the equation becomes:
\be
\big(D^3+\mu_1E_4(\tau)D+\mu_2E_6(\tau)\big)\chi=0
\ee
While this equation was first studied long ago in \cite{Mathur:1988gt}, all its admissible solutions with $c\leq 96$ have been classified only recently \cite{Das:2021uvd} and we will make use of some of these below (all these relevant solutions are given in table \ref{t6} in Sec. \ref{a3}).

Next we summarise some results about the coset construction. In \cite{Gaberdiel:2016zke} the following relation between the central charge $c$ and conformal dimensions $h_i$ of the characters $\chi_i$, and the corresponding quantities $\tc,\tilh_i$ of the coset dual $\tchi_i$ was derived:
\be
\ell+\tl = n^2+\left(\frac{c+\tc}{4}-1\right)n-6\sum_{i=1}^{n-1}(h_i+\tilh_i) 
\ee
We will be interested in the case $n=3$. Also, note that $c+\tc$ must be a multiple of 8, so we write it as $8N$ where $N$ is an integer $\ge 1$ \footnote{In \cite{Gaberdiel:2016zke} the convention was to write $c+\tc=24N$ where $N$ is a multiple of $\frac13$.}. Also, since the right hand side of the bilinear relation is a character with integer dimensions, we must have $h_i+\tilh_i=n_i$, an integer $\ge 1$, for each $i$. Thus, for $n=3$ the relation can be written:
\be
\ell+\tl= 6\Big(N+1-\sum_{i=1}^{2}n_i\Big)
\label{neq3reln}
\ee

Let us comment on the significance of the integers $n_i$. These denote the new chiral currents, of spin $n_i$, created by ``fusing'' the primaries of two RCFTs $\cC,\tcC$ via a bilinear relation into a meromorphic theory $\cH$. Whenever each $n_i\ge 2$, we have a special situation where the only new currents in $\cH$, relative to $\cC\oplus \tcC$, have spin $\ge 2$. In such cases the Kac-Moody algebra of $\cH$ is necessarily the direct sum of that of $\cC,\tcC$. On the other hand if any $n_i=1$, we have new Kac-Moody currents in $\cH$ that were not present in the coset pair. Viewed from the converse perspective starting with the meromorphic theory, the first case $n_i\ge 2$ arises when the denominator theory $\cC$ is an affine theory for one (or more) of the simple factors in $\cH$. In that case the coset procedure just deletes the factor(s) and the remaining ones make up the Kac-Moody algebra of $\tcC$. But in the second case, the algebra of $\cC$ undergoes a non-trivial embedding in one of the simple factors of $\cH$. This ``Higgses'' that factor down to a subalgebra given by the commutant, and some currents are lost in the process. Only the first case $n_i\ge 2$  will be of relevance in the present work, while the other case involving  embeddings will be discussed in \cite{DGM:upcoming,MR:upcoming}.

From now on we focus on coset dual pairs of characters with $n=3$ and $\ell,{\tilde \ell}=0,0$. From the above relation it follows that $n_1+n_2=N+1$. While the classification of admissible $(3,0)$ characters has not been completed, significant progress has been made in \cite{Mathur:1988gt,Tuite:2008pt,Franc:2019, Gaberdiel:2016zke, Das:2021uvd}.
We will make use of the results in \cite{Das:2021uvd}, which subsumes the output of most of the previous works and provides a complete set of admissible three-character sets with $c\le 96$ \footnote{In an upcoming work \cite{DGM:upcoming} we re-examine this partial classification of admissible characters and attempt to identify which of them can be shown to exist as CFTs as well as which ones can be shown not to correspond to any CFT.}.
 
In the notation of \cite{Das:2021uvd}, to which we refer the reader for more details, the admissible character sets fall into five categories, labelled $\mathbf{I},\mathbf{II},\cdots,\mathbf{V}$. Of these, all entries in categories $\mathbf{I},\mathbf{II},\mathbf{IV}$ have already been identified as CFTs in \cite{Das:2021uvd} \footnote{with a few exceptions that correspond to generalisations of CFTs such as Intermediate Vertex Operator Algebras (IVOA) \cite{Kawasetsu:2014}.}. Among these are characters that were previously identified as CFTs in \cite{Gaberdiel:2016zke} and which will play a role in the present work. We will label them GHM$_{D}$ where the subscript $D$ denotes the dimension of their Kac-Moody algebra (see all GHM solutions listed in table \ref{t6} in Sec. \ref{a3}). We will also make use of characters of type $\mathbf{III}$ and $\mathbf{V}$ which were not identified with CFTs in \cite{Das:2021uvd}. These will be studied in complete detail in work to appear \cite{DGM:upcoming}. The list of relevant ${\bf III}$ and ${\bf V}$ solutions can be found in table \ref{t6} of Appendix \ref{a3}. 



\section{Constructing new meromorphic CFTs}

\subsection{$(3,0)$ cosets from $c=24$ meromorphic theories}

We start by using the coset construction with $\cH$ being one of the meromorphic theories in Schellekens' list, to identify 22 sets of admissible characters as CFTs. Table \ref{t0} shows coset pairings of characters $\chi_i,{\tilde\chi}_i$ to make one of these meromorphic theories. The purpose of this exercise is to identify the theories $\tilde{\mathcal{C}}$ that we will use later on. 

The entries in the table are as follows. The first column is a serial number labelling the 22 cases of interest. The next four columns tell us the properties of an affine (or minimal) model $\cC$ that we use as the denominator in the coset relation. Respectively, they provide the central charge, conformal dimensions, dimension $m_1$ of the Kac-Moody algebra, and the Kac-Moody algebra itself. The next four columns provide the same properties for a coset theory $\tcC$ that combines with $\cC$ in a bilinear relation that has been verified. The last four columns tell us the Kac-Moody algebra of the meromorphic Schellekens theory, the integers $(d_1,d_2)$ appearing in the bilinear relation \eref{bilinrel}, the integer ${\cal N}$ that specifies the character via $\chi^\cH(\tau)=j(\tau)+{\cal N}$, and the serial number of the corresponding theory in the table of \cite{Schellekens:1992db}. Note that the dimension of the Kac-Moody algebra of the meromorphic theory is ${\cal N}+744$.

As can be seen in the Table, rows $4-9,13-17,22$ are cases where $\tilde{\mathcal{C}}$ is of GHM type \cite{Gaberdiel:2016zke}. Among the rest, rows $1,2,10-12,18-21$ were missed in \cite{Gaberdiel:2016zke} for various reasons -- for example that reference did not consider coset duals where $\cC$ is a tensor product of two identical two-character theories. Note that despite this, $\tcC$ is {\em not} a tensor product of simpler theories. Finally, row 3 is the Ising-Baby Monster pairing implicit in \cite{Hoehn:Baby8} and discussed in the present context in \cite{Hampapura:2016mmz}. This is the only known case of a $c=24$ coset where one or both (in this case, both) of the entries have no Kac-Moody algebra. The coset relation must then be understood in the more general sense mentioned below \eref{bilinrel}, and the corresponding meromorphic theory is the Monster CFT. We will soon see that for $c\ge 32$, the Baby Monster theory features in coset pairings with affine theories. 


\setlength\LTleft{-50pt}
\setlength\LTright{0pt}
\begin{longtable}{l||cccc||cccc||cccc}
\caption{Coset relations for $c^\mathcal{H}=24$ with $(n_1,n_2)=(2,2)$. We identify $\mathcal{M}(4,3)\cong B_{0,1}$, $A_{1,2}\cong B_{1,1}$, $C_{2,1}\cong B_{2,1}$ and $\text{BM}$ $\equiv$ Baby Monster. We further identify $U(1)\cong D_{1,1}$, $A_{1,1}^{\oplus 2}\cong D_{2,1}$ and $A_{3,1}\cong D_{3,1}$.}
\label{t0}\\
\hline
\hline
\makebox[0pt][l]{\fboxsep0pt\colorbox{Mywhite} {\strut\hspace*{1.2\linewidth}}}
\# & $c$ & $(h_1,h_2)$ & $m_1$ & $\mathcal{C}$ & $\Tilde{c}$ & $(\Tilde{h}_1,\Tilde{h}_2)$ & $\tilde{m}_1$ & $\Tilde{\mathcal{C}}$ & Affine & \tiny{$(d_1, d_2)$} & $\mathcal{N}$ & Sch. \\
\makebox[0pt][l]{\fboxsep0pt\colorbox{Mywhite} {\strut\hspace*{1.2\linewidth}}}
 &  &  &  &  &  &  &  &  & algebra &  &  & \# \\
\hline
\hline
\makebox[0pt][l]{\fboxsep0pt\colorbox{Mywhite} {\strut\hspace*{1.2\linewidth}}}
1. & $12$ & $(\frac{1}{2},\frac{3}{2})$ & $276$ &$D_{12,1}$ & $12$ & $(\frac{3}{2},\frac{1}{2})$ & $276$ & $D_{12,1}$ & $D_{12,1}^{\oplus 2}$ & \tiny{$2$} & 552 & 66 \\
\makebox[0pt][l]{\fboxsep0pt\colorbox{Mygrey} {\strut\hspace*{1.2\linewidth}}}
2. & $12$ & $(\frac{2}{3},\frac{4}{3})$ & $156$ & $E_{6,1}^{\oplus 2}$ & $12$ & $(\frac{4}{3},\frac{2}{3})$ & $156$ & $E_{6,1}^{\oplus 2}$ & $E_{6,1}^{\oplus 4}$ & \tiny{$8$} & 312 & 58 \\
\makebox[0pt][l]{\fboxsep0pt\colorbox{Mywhite} {\strut\hspace*{1.2\linewidth}}}
3. & $\frac{1}{2}$ & $(\frac{1}{2},\frac{1}{16})$ & $0$ & $\mathcal{M}(4,3)$ & $\frac{47}{2}$ & $(\frac{3}{2},\frac{31}{16})$ & $0$ & $\text{BM}$ & --- & \tiny{$(1,1)$} & 0 & 0 \\
\makebox[0pt][l]{\fboxsep0pt\colorbox{Mygrey} {\strut\hspace*{1.2\linewidth}}}
4. & $\frac{3}{2}$ & $(\frac{1}{2},\frac{3}{16})$ & $3$ & $A_{1,2}$ & $\frac{45}{2}$ & $(\frac{3}{2},\frac{29}{16})$ & $45$ & $\text{GHM}_{45}$ & $A_{1,2}^{\oplus 16}$ & \tiny{$(1,1024)$} & 48 & 5 \\
\makebox[0pt][l]{\fboxsep0pt\colorbox{Mygrey} {\strut\hspace*{1.2\linewidth}}}
 &  &  &  &  &  &  &  &  & $A_{3,4}^{\oplus 3}A_{1,2}$ &  &  & 7 \\
\makebox[0pt][l]{\fboxsep0pt\colorbox{Mygrey} {\strut\hspace*{1.2\linewidth}}}
 &  &  &  &  &  &  &  &  & $A_{5,6}C_{2,3}A_{1,2}$ &  &  & 8 \\
\makebox[0pt][l]{\fboxsep0pt\colorbox{Mygrey} {\strut\hspace*{1.2\linewidth}}}
 &  &  &  &  &  &  &  &  & $D_{5,8}A_{1,2}$ &  &  & 10 \\
\makebox[0pt][l]{\fboxsep0pt\colorbox{Mywhite} {\strut\hspace*{1.2\linewidth}}}
5. & $\frac{5}{2}$ & $(\frac{1}{2},\frac{5}{16})$ & $10$ & $C_{2,1}$ & $\frac{43}{2}$ & $(\frac{3}{2},\frac{27}{16})$ & $86$ & $\text{GHM}_{86}$ & $D_{4,2}^{\oplus 2}C_{2,1}^{\oplus 4}$ & \tiny{$(1,512)$} & 96 & 25 \\
\makebox[0pt][l]{\fboxsep0pt\colorbox{Mywhite} {\strut\hspace*{1.2\linewidth}}}
 &  &  &  &  &  &  &  &  & $A_{5,2}^{\oplus 2}A_{2,1}^{\oplus 2}C_{2,1}$ &  &  & 26 \\
\makebox[0pt][l]{\fboxsep0pt\colorbox{Mywhite} {\strut\hspace*{1.2\linewidth}}}
 &  &  &  &  &  &  &  &  & $E_{6,4}A_{2,1}C_{2,1}$ &  &  & 28 \\ 
\makebox[0pt][l]{\fboxsep0pt\colorbox{Mygrey} {\strut\hspace*{1.2\linewidth}}}
6. & $\frac{7}{2}$ & $(\frac{1}{2},\frac{7}{16})$ & $21$ & $B_{3,1}$ & $\frac{41}{2}$ & $(\frac{3}{2},\frac{25}{16})$ & $123$ & $\text{GHM}_{123}$ & $B_{3,1}D_{6,2}C_{4,1}B_{3,1}$ & \tiny{$(1,256)$} & 144 & 39 \\
\makebox[0pt][l]{\fboxsep0pt\colorbox{Mygrey} {\strut\hspace*{1.2\linewidth}}}
 &  &  &  &  &  &  &  &  & $B_{3,1}A_{9,2}A_{4,1}$ &  &  & 40 \\
\makebox[0pt][l]{\fboxsep0pt\colorbox{Mywhite} {\strut\hspace*{1.2\linewidth}}}
7. & $\frac{9}{2}$ & $(\frac{1}{2},\frac{9}{16})$ & $36$ & $B_{4,1}$ & $\frac{39}{2}$ & $(\frac{3}{2},\frac{23}{16})$ & $156$ & $\text{GHM}_{156}$ & $D_{8,2}B_{4,1}^{\oplus 2}$ & \tiny{$(1,128)$} & 192 & 47 \\
\makebox[0pt][l]{\fboxsep0pt\colorbox{Mywhite} {\strut\hspace*{1.2\linewidth}}}
 &  &  &  &  &  &  &  &  & $B_{4,1}C_{6,1}^{\oplus 2}$ &  &  & 48 \\
\makebox[0pt][l]{\fboxsep0pt\colorbox{Mygrey} {\strut\hspace*{1.2\linewidth}}}
8. & $\frac{11}{2}$ & $(\frac{1}{2},\frac{11}{16})$ & $55$ & $B_{5,1}$ & $\frac{37}{2}$ & $(\frac{3}{2},\frac{21}{16})$ & $185$ & $\text{GHM}_{185}$ & $B_{5,1}E_{7,2}F_{4,1}$ & \tiny{$(1,1)$} & 240 & 53 \\
\makebox[0pt][l]{\fboxsep0pt\colorbox{Mywhite} {\strut\hspace*{1.2\linewidth}}}
9. & $\frac{13}{2}$ & $(\frac{1}{2},\frac{13}{16})$ & $78$ & $B_{6,1}$ & $\frac{35}{2}$ & $(\frac{3}{2},\frac{19}{16})$ & $210$ & $\text{GHM}_{210}$ & $B_{6,1}C_{10,1}$ & \tiny{$(1,32)$} & 288 & 56 \\
\makebox[0pt][l]{\fboxsep0pt\colorbox{Mygrey} {\strut\hspace*{1.2\linewidth}}}
10. & $\frac{17}{2}$ & $(\frac{1}{2},\frac{17}{16})$ & $136$ & $B_{8,1}$ & $\frac{31}{2}$ & $(\frac{3}{2},\frac{15}{16})$ & $248$ & $E_{8,2}$ & $B_{8,1}E_{8,2}$ & \tiny{$(1,1)$} & 384 & 62 \\
\makebox[0pt][l]{\fboxsep0pt\colorbox{Mywhite} {\strut\hspace*{1.2\linewidth}}}
11. & $1$ & $(\frac{1}{2},\frac{1}{8})$ & $1$ & $U(1)$ & $23$ & $(\frac{3}{2},\frac{15}{8})$ & $23$ & ${\bf III_{50}}$ & $U(1)^{\oplus 24}$ & \tiny{$(8,4096)$} & 24 & 1 \\
\makebox[0pt][l]{\fboxsep0pt\colorbox{Mygrey} {\strut\hspace*{1.2\linewidth}}}
12. & $2$ & $(\frac{1}{2},\frac{1}{4})$ & $6$ & $A_{1,1}^{\oplus 2}$ & $22$ & $(\frac{3}{2},\frac{7}{4})$ & $66$ & ${\bf III_{45}}$ &$A_{1,1}^{\oplus 24}$ & \tiny{$(64,4096)$} & 72 & 15 \\
\makebox[0pt][l]{\fboxsep0pt\colorbox{Mygrey} {\strut\hspace*{1.2\linewidth}}}
 &  &  &  &  &  &  &  &  & $A_{3,2}^{\oplus 4} A_{1,1}^{\oplus 4}$ &  &  & 16 \\
\makebox[0pt][l]{\fboxsep0pt\colorbox{Mygrey} {\strut\hspace*{1.2\linewidth}}}
 &  &  &  &  &  &  &  &  & $A_{5,3}D_{4,3}A_{1,1}^{\oplus 3}$ &  &  & 17 \\
\makebox[0pt][l]{\fboxsep0pt\colorbox{Mygrey} {\strut\hspace*{1.2\linewidth}}}
 &  &  &  &  &  &  &  &  & $A_{7,4}A_{1,1}^{\oplus 3}$ &  &  & 18 \\
\makebox[0pt][l]{\fboxsep0pt\colorbox{Mygrey} {\strut\hspace*{1.2\linewidth}}}
 &  &  &  &  &  &  &  &  & $D_{5,4}C_{3,2}A_{1,1}^{\oplus 2}$ &  &  & 19 \\
\makebox[0pt][l]{\fboxsep0pt\colorbox{Mygrey} {\strut\hspace*{1.2\linewidth}}}
 &  &  &  &  &  &  &  &  & $D_{6,5}A_{1,1}^{\oplus 2}$ &  &  & 20 \\
\makebox[0pt][l]{\fboxsep0pt\colorbox{Mywhite} {\strut\hspace*{1.2\linewidth}}}
13. & $3$ & $(\frac{1}{2},\frac{3}{8})$ & $15$ & $A_{3,1}$ & $21$ & $(\frac{3}{2},\frac{13}{8})$ & $105$ & $\text{GHM}_{105}$ & $A_{3,1}^{\oplus 8}$ & \tiny{$(8,1024)$} & 120 & 30 \\
\makebox[0pt][l]{\fboxsep0pt\colorbox{Mywhite} {\strut\hspace*{1.2\linewidth}}}
 &  &  &  &  &  &  &  &  & $D_{5,2}^{\oplus 2}A_{3,1}^{\oplus 2}$ &  &  & 31 \\
\makebox[0pt][l]{\fboxsep0pt\colorbox{Mywhite} {\strut\hspace*{1.2\linewidth}}}
 &  &  &  &  &  &  &  &  & $A_{7,2}C_{3,1}^{\oplus 2}A_{3,1}$ &  &  & 33 \\
\makebox[0pt][l]{\fboxsep0pt\colorbox{Mywhite} {\strut\hspace*{1.2\linewidth}}}
 &  &  &  &  &  &  &  &  & $D_{7,3}G_{2,1}A_{3,1}$ &  &  & 34 \\
\makebox[0pt][l]{\fboxsep0pt\colorbox{Mywhite} {\strut\hspace*{1.2\linewidth}}}
 &  &  &  &  &  &  &  &  & $C_{7,2}A_{3,1}$ &  &  & 35 \\  
\makebox[0pt][l]{\fboxsep0pt\colorbox{Mygrey} {\strut\hspace*{1.2\linewidth}}}
14. & $5$ & $(\frac{1}{2},\frac{5}{8})$ & $45$ & $D_{5,1}$ & $19$ & $(\frac{3}{2},\frac{11}{8})$ & $171$ & $\text{GHM}_{171}$ & $D_{5,1}^{\oplus 2} \, A_{7,1}^{\oplus 2}$ & \tiny{$(8,256)$} & 216 & 49 \\
\makebox[0pt][l]{\fboxsep0pt\colorbox{Mywhite} {\strut\hspace*{1.2\linewidth}}}
15. & $6$ & $(\frac{1}{2},\frac{3}{4})$ & $66$ & $D_{6,1}$ & $18$ & $(\frac{3}{2},\frac{5}{4})$ & $198$ & $\text{GHM}_{198}$ & $D_{6,1}^{\oplus 4}$ & \tiny{$(64,256)$} & 264 & 54 \\
\makebox[0pt][l]{\fboxsep0pt\colorbox{Mywhite} {\strut\hspace*{1.2\linewidth}}}
 &  &  &  &  &  &  &  &  & $D_{6,1} \, A_{9,1}^{\oplus 2}$ &  &  & 55 \\
\makebox[0pt][l]{\fboxsep0pt\colorbox{Mygrey} {\strut\hspace*{1.2\linewidth}}}
16. & $7$ & $(\frac{1}{2},\frac{7}{8})$ & $91$ & $D_{7,1}$ & $17$ & $(\frac{3}{2},\frac{9}{8})$ & $221$ & $\text{GHM}_{221}$ & $D_{7,1} \, A_{11,1} \, E_{6,1}$ & \tiny{$(8,64)$} & 312 & 59 \\ 
\makebox[0pt][l]{\fboxsep0pt\colorbox{Mywhite} {\strut\hspace*{1.2\linewidth}}}
17. & $9$ & $(\frac{1}{2},\frac{9}{8})$ & $153$ & $D_{9,1}$ & $15$ & $(\frac{3}{2},\frac{7}{8})$ & $255$ & $\text{GHM}_{255}$ & $D_{9,1} \, A_{15,1}$ & \tiny{$(8,16)$} & 408 & 63 \\ 
\makebox[0pt][l]{\fboxsep0pt\colorbox{Mygrey} {\strut\hspace*{1.2\linewidth}}}
18. & $10$ & $(\frac{1}{2},\frac{5}{4})$ & $190$ & $D_{10,1}$ & $14$ & $(\frac{3}{2},\frac{3}{4})$ & $266$ & $E_{7,1}^{\oplus 2}$ & $D_{10,1}E_{7,1}^{\oplus 2}$ & \tiny{$(1,2)$} & 456 & 64 \\
\makebox[0pt][l]{\fboxsep0pt\colorbox{Mywhite} {\strut\hspace*{1.2\linewidth}}}
19. & $4$ & $(\frac{1}{3},\frac{2}{3})$ & $16$ & $A_{2,1}^{\oplus 2}$ & $20$ & $(\frac{5}{3},\frac{4}{3})$ & $80$ & ${\bf V_{39}}$ & $A_{2,1}^{\oplus 12}$ & \tiny{$(8748,972)$} & 96 & 24 \\
\makebox[0pt][l]{\fboxsep0pt\colorbox{Mywhite} {\strut\hspace*{1.2\linewidth}}}
 &  &  &  &  &  &  &  &  & $A_{5,2}^{\oplus 2}C_{2,1}A_{2,1}^{\oplus 2}$ &  &  & 26 \\
\makebox[0pt][l]{\fboxsep0pt\colorbox{Mywhite} {\strut\hspace*{1.2\linewidth}}}
 &  &  &  &  &  &  &  &  & $A_{8,3}A_{2,1}^{\oplus 2}$ &  &  & 27 \\
\makebox[0pt][l]{\fboxsep0pt\colorbox{Mygrey} {\strut\hspace*{1.2\linewidth}}}
20. & $\frac{28}{5}$ & $(\frac{2}{5},\frac{4}{5})$ & $28$ & $G_{2,1}^{\oplus 2}$ & $\frac{92}{5}$ & $(\frac{8}{5},\frac{6}{5})$ & $92$ & ${\bf III_{37}}$ & $E_{6,3}G_{2,1}^{\oplus 3}$ & \tiny{$(50,1)$} & 120 & 32 \\ 
\makebox[0pt][l]{\fboxsep0pt\colorbox{Mywhite} {\strut\hspace*{1.2\linewidth}}}
21. & $\frac{52}{5}$ & $(\frac{3}{5},\frac{6}{5})$ & $104$ & $F_{4,1}^{\oplus 2}$ & $\frac{68}{5}$ & $(\frac{7}{5},\frac{4}{5})$ & $136$ & ${\bf III_{22}}$ & $C_{8,1}F_{4,1}^{\oplus 2}$ & \tiny{$(50,1)$} & 240 & 52 \\
\makebox[0pt][l]{\fboxsep0pt\colorbox{Mygrey} {\strut\hspace*{1.2\linewidth}}}
22. & $4$ & $(\frac{2}{5},\frac{3}{5})$ & $24$ & $A_{4,1}$ & $20$ & $(\frac{8}{5},\frac{7}{5})$ & $120$ & $\text{GHM}_{120}$ & $A_{4,1}^{\oplus 6}$ & \tiny{$(1250,1250)$} & 144 & 37 \\
\makebox[0pt][l]{\fboxsep0pt\colorbox{Mygrey} {\strut\hspace*{1.2\linewidth}}}
 &  &  &  &  &  &  &  &  & $A_{9,2}B_{3,1}A_{4,1}$ &  &  & 40 \\
\hline
\hline
\end{longtable} 
Table \ref{t0} includes a few self-dual pairs. In these cases we have: $\chi_0=\Tilde{\chi}_0$, $\chi_1=\Tilde{\chi}_2$ and $\chi_2=\Tilde{\chi}_1$. Hence Eq.(\ref{bilinrel}) (for three characters) becomes:
    \begin{align*}
        \chi_0^\mathcal{H} = \chi_0^2 + d_1 \, \chi_1\chi_2 + d_2 \, \chi_2\chi_1 = \chi_0^2 + (d_1+d_2) \, \chi_1\chi_2 = \chi_0^2 + d_3 \, \chi_1\chi_2.
    \end{align*}
Hence, in the $(d_1,d_2)$ column we just have one entry $d_3=d_1+d_2$ for self-dual pairs. We follow this convention for all tables whenever there are self-dual pairs involved in a bilinear relation.



\subsection{New meromorphic theories at $c\ge 32$: infinite families}

The next step is to combine the theories labelled $\tcC$ in Table \ref{t0} with suitable infinite series of affine theories to make modular invariants with central charge $\ge 32$. One set of results is exhibited in Table \ref{t1}, where 34 infinite families of coset pairs are described. The theories labelled $\tcC$ are all taken from Table \ref{t0}. However in each case the corresponding theory labelled $\cC$ in Table \ref{t0} has been replaced by a new affine theory $\cC$ labelled by an arbitrary integer parameter $m$.  The associated character is exhibited in the last column of the table (see below for an explanation of the notation). The $m=0$ case is the original one in Table \ref{t0}. The integers $(n_1,n_2)$ are, in all cases, given by $(2,m+2)$, verifying \eref{neq3reln} between a pair of three-characters with vanishing Wronskian index.

In this table, both $\cC$ and $\tcC$ correspond to known theories. Hence, from the coset pairing we conclude that the modular invariant obtained by combining them in a bilinear relation also corresponds to a genuine CFT. This is a meromorphic CFT that, in most cases, has to be of non-lattice type since it either involves non-simply-laced algebras or levels greater than 1, or both.


\setlength\LTleft{-30pt}
\setlength\LTright{0pt}
\begin{longtable}{l||cccc||cccc||c} 
\caption{Coset relations for $c^\mathcal{H}=8(m+3)$ with $(n_1,n_2)=(2,m+2)$.}
\label{t1}\\
\hline
\hline
\makebox[0pt][l]{\fboxsep0pt\colorbox{Mywhite} {\strut\hspace*{1.12\linewidth}}}
\# & $c$ & \tiny{$(h_1,h_2)$} & \tiny{$m_1$} & $\mathcal{C}$ & $\Tilde{c}$ & \tiny{$(\Tilde{h}_1,\Tilde{h}_2)$} & \tiny{$\Tilde{m}_1$} & $\Tilde{\mathcal{C}}$ & Sch. \# \\
\makebox[0pt][l]{\fboxsep0pt\colorbox{Mywhite} {\strut\hspace*{1.12\linewidth}}}
  &  &  &  &  &  &  &  &  & ($m=0$) \\
\hline
\hline
\makebox[0pt][l]{\fboxsep0pt\colorbox{Mywhite} {\strut\hspace*{1.12\linewidth}}}
1. & $\frac{16m+1}{2}$ & \tiny{$\left(\frac{1}{2},\frac{16m+1}{16}\right)$} & \tiny{128$m^2$ + 8$m$} & $B_{8m,1}$ & $\frac{47}{2}$ & \tiny{$\left(\frac{3}{2},\frac{31}{16}\right)$} & \tiny{$0$} & $\text{BM}$ & 0 \\
\makebox[0pt][l]{\fboxsep0pt\colorbox{Mygrey} {\strut\hspace*{1.12\linewidth}}}
2. & $\frac{16m+3}{2}$ & \tiny{$\left(\frac{1}{2},\frac{16m+3}{16}\right)$} & \tiny{128$m^2$ + 40$m$ + 3} & $B_{8m+1,1}$ & $\frac{45}{2}$ & \tiny{$\left(\frac{3}{2},\frac{29}{16}\right)$} & \tiny{$45$} & $\mathcal{E}_3[A_{1,2}^{\oplus 15}]$ & 5 \\
\makebox[0pt][l]{\fboxsep0pt\colorbox{Mygrey} {\strut\hspace*{1.12\linewidth}}}
  &  &  &  &  &  &  &  & $\mathcal{E}_3[A_{3,4}^{\oplus 3}]$ & 7 \\
\makebox[0pt][l]{\fboxsep0pt\colorbox{Mygrey} {\strut\hspace*{1.12\linewidth}}}
  &  &  &  &  &  &  &  & $\mathcal{E}_3[A_{5,6}C_{2,3}]$ & 8 \\
\makebox[0pt][l]{\fboxsep0pt\colorbox{Mygrey} {\strut\hspace*{1.12\linewidth}}}
  &  &  &  &  &  &  &  & $\mathcal{E}_3[D_{5,8}]$ & 10 \\ 
\makebox[0pt][l]{\fboxsep0pt\colorbox{Mywhite} {\strut\hspace*{1.12\linewidth}}}
3. & $\frac{16m+5}{2}$ & \tiny{$\left(\frac{1}{2},\frac{16m+5}{16}\right)$} & \tiny{128$m^2$ + 72$m$ + 10} & $B_{8m+2,1}$ & $\frac{43}{2}$ & \tiny{$\left(\frac{3}{2},\frac{27}{16}\right)$} & \tiny{$45$} & $\mathcal{E}_3[D_{4,2}^{\oplus 2}C_{2,1}^{\oplus 3}]$ & 25 \\
\makebox[0pt][l]{\fboxsep0pt\colorbox{Mywhite} {\strut\hspace*{1.12\linewidth}}}
  &  &  &  &  &  &  &  & $\mathcal{E}_3[A_{5,2}^{\oplus 2}A_{2,1}^{\oplus 2}]$ & 26 \\
\makebox[0pt][l]{\fboxsep0pt\colorbox{Mywhite} {\strut\hspace*{1.12\linewidth}}}
  &  &  &  &  &  &  &  & $\mathcal{E}_3[E_{6,4}A_{2,1}]$ & 28 \\
\makebox[0pt][l]{\fboxsep0pt\colorbox{Mygrey} {\strut\hspace*{1.12\linewidth}}}
4. & $\frac{16m+7}{2}$ & \tiny{$\left(\frac{1}{2},\frac{16m+7}{16}\right)$} & \tiny{128$m^2$ + 104$m$ + 21} & $B_{8m+3,1}$ & $\frac{41}{2}$ & \tiny{$\left(\frac{3}{2},\frac{25}{16}\right)$} & \tiny{$123$} & $\mathcal{E}_3[D_{6,2}C_{4,1}B_{3,1}]$ & 39 \\
\makebox[0pt][l]{\fboxsep0pt\colorbox{Mygrey} {\strut\hspace*{1.12\linewidth}}}
  &  &  &  &  &  &  &  & $\mathcal{E}_3[A_{9,2}A_{4,1}]$ & 40 \\
\makebox[0pt][l]{\fboxsep0pt\colorbox{Mywhite} {\strut\hspace*{1.12\linewidth}}}
5. & $\frac{16m+9}{2}$ & \tiny{$\left(\frac{1}{2},\frac{16m+9}{16}\right)$} & \tiny{128$m^2$ + 136$m$ + 36} & $B_{8m+4,1}$ & $\frac{39}{2}$ & \tiny{$\left(\frac{3}{2},\frac{23}{16}\right)$} & \tiny{$156$} & $\mathcal{E}_3[D_{8,2}B_{4,1}]$ & 47 \\
\makebox[0pt][l]{\fboxsep0pt\colorbox{Mywhite} {\strut\hspace*{1.12\linewidth}}}
  &  &  &  &  &  &  &  & $\mathcal{E}_3[C_{6,1}^{\oplus 2}]$ & 48 \\  
\makebox[0pt][l]{\fboxsep0pt\colorbox{Mygrey} {\strut\hspace*{1.12\linewidth}}}
6. & $\frac{16m+11}{2}$ & \tiny{$\left(\frac{1}{2},\frac{16m+11}{16}\right)$} & \tiny{128$m^2$ + 168$m$ + 55} & $B_{8m+5,1}$ & $\frac{37}{2}$ & \tiny{$\left(\frac{3}{2},\frac{21}{16}\right)$} & \tiny{$185$} & $\mathcal{E}_3[E_{7,2}F_{4,1}]$ & 53 \\  
\makebox[0pt][l]{\fboxsep0pt\colorbox{Mywhite} {\strut\hspace*{1.12\linewidth}}}
7. & $\frac{16m+13}{2}$ & \tiny{$\left(\frac{1}{2},\frac{16m+13}{16}\right)$} & \tiny{128$m^2$ + 200$m$ + 78} & $B_{8m+6,1}$ & $\frac{35}{2}$ & \tiny{$\left(\frac{3}{2},\frac{19}{16}\right)$} & \tiny{$210$} & $\mathcal{E}_3[C_{10,1}]$ & 56 \\ 
\makebox[0pt][l]{\fboxsep0pt\colorbox{Mygrey} {\strut\hspace*{1.12\linewidth}}}
8. & $\frac{16m+17}{2}$ & \tiny{$\left(\frac{1}{2},\frac{16m+17}{16}\right)$} & \tiny{128$m^2$ + 264$m$ + 136} & $B_{8m+8,1}$ & $\frac{31}{2}$ & \tiny{$\left(\frac{3}{2},\frac{15}{16}\right)$} & \tiny{$248$} & $E_{8,2}$ & 62 \\  
\makebox[0pt][l]{\fboxsep0pt\colorbox{Mywhite} {\strut\hspace*{1.12\linewidth}}}
9. & $8m+1$ & \tiny{$\left(\frac{1}{2},\frac{8m+1}{8}\right)$} & \tiny{128$m^2$ + 24$m$ + 1} & $D_{8m+1,1}$ & $23$ & \tiny{$\left(\frac{3}{2},\frac{15}{8}\right)$} & \tiny{$23$} & $\mathcal{E}_3[D_{1,1}^{\oplus 23}]$ & 1 \\
\makebox[0pt][l]{\fboxsep0pt\colorbox{Mygrey} {\strut\hspace*{1.12\linewidth}}}
10. & $8m+2$ & \tiny{$\left(\frac{1}{2},\frac{8m+2}{8}\right)$} & \tiny{128$m^2$ + 56$m$ + 6} & $D_{8m+2,1}$ & $22$ & \tiny{$\left(\frac{3}{2},\frac{7}{4}\right)$} & \tiny{$66$} & $\mathcal{E}_3[A_{1,1}^{\oplus 22}]$ & 15 \\
\makebox[0pt][l]{\fboxsep0pt\colorbox{Mygrey} {\strut\hspace*{1.12\linewidth}}}
  &  &  &  &  &  &  &  & $\mathcal{E}_3[A_{3,2}^{\oplus 4}A_{1,1}^{\oplus 2}]$ & 16 \\
\makebox[0pt][l]{\fboxsep0pt\colorbox{Mygrey} {\strut\hspace*{1.12\linewidth}}}
  &  &  &  &  &  &  &  & $\mathcal{E}_3[A_{5,3}D_{4,3}A_{1,1}]$ & 17 \\
\makebox[0pt][l]{\fboxsep0pt\colorbox{Mygrey} {\strut\hspace*{1.12\linewidth}}}
  &  &  &  &  &  &  &  & $\mathcal{E}_3[A_{7,4}A_{1,1}]$ & 18  \\ 
\makebox[0pt][l]{\fboxsep0pt\colorbox{Mygrey} {\strut\hspace*{1.12\linewidth}}}
  &  &  &  &  &  &  &  & $\mathcal{E}_3[D_{5,4}C_{3,2}]$ & 19  \\ 
\makebox[0pt][l]{\fboxsep0pt\colorbox{Mygrey} {\strut\hspace*{1.12\linewidth}}}
  &  &  &  &  &  &  &  & $\mathcal{E}_3[D_{6,5}]$ & 20  \\   
\makebox[0pt][l]{\fboxsep0pt\colorbox{Mywhite} {\strut\hspace*{1.12\linewidth}}}
11. & $8m+3$ & \tiny{$\left(\frac{1}{2},\frac{8m+3}{8}\right)$} & \tiny{128$m^2$ + 88$m$ + 15} & $D_{8m+3,1}$ & $21$ & \tiny{$\left(\frac{3}{2},\frac{13}{8}\right)$} & \tiny{$105$} & $\mathcal{E}_3[A_{3,1}^{\oplus 7}]$ & 30 \\
\makebox[0pt][l]{\fboxsep0pt\colorbox{Mywhite} {\strut\hspace*{1.12\linewidth}}}
  &  &  &  &  &  &  &  & $\mathcal{E}_3[D_{5,2}^{\oplus 2}A_{3,1}]$ & 31   \\
\makebox[0pt][l]{\fboxsep0pt\colorbox{Mywhite} {\strut\hspace*{1.12\linewidth}}}
  &  &  &  &  &  &  &  & $\mathcal{E}_3[A_{7,2}C_{3,1}^{\oplus 2}]$ & 33  \\
\makebox[0pt][l]{\fboxsep0pt\colorbox{Mywhite} {\strut\hspace*{1.12\linewidth}}}
  &  &  &  &  &  &  &  & $\mathcal{E}_3[D_{7,3}G_{2,1}]$ & 34  \\ 
\makebox[0pt][l]{\fboxsep0pt\colorbox{Mywhite} {\strut\hspace*{1.12\linewidth}}}
  &  &  &  &  &  &  &  & $\mathcal{E}_3[C_{7,2}]$ & 35  \\     
\makebox[0pt][l]{\fboxsep0pt\colorbox{Mygrey} {\strut\hspace*{1.12\linewidth}}}
12. & $8m+5$ & \tiny{$\left(\frac{1}{2},\frac{8m+5}{8}\right)$} & \tiny{128$m^2$ + 152$m$ + 45} & $D_{8m+5,1}$ & $19$ & \tiny{$\left(\frac{3}{2},\frac{11}{8}\right)$} & \tiny{$171$} & $\mathcal{E}_3[D_{5,1}A_{7,1}^{\oplus 2}]$ & 49 \\
\makebox[0pt][l]{\fboxsep0pt\colorbox{Mywhite} {\strut\hspace*{1.12\linewidth}}}
13. & $8m+6$ & \tiny{$\left(\frac{1}{2},\frac{8m+6}{8}\right)$} & \tiny{128$m^2$ + 184$m$ + 66} & $D_{8m+6,1}$ & $18$ & \tiny{$\left(\frac{3}{2},\frac{5}{4}\right)$} & \tiny{$198$} & $\mathcal{E}_3[D_{6,1}^{\oplus 3}]$ & 54 \\
\makebox[0pt][l]{\fboxsep0pt\colorbox{Mywhite} {\strut\hspace*{1.12\linewidth}}}
  &  &  &  &  &  &  &  & $\mathcal{E}_3[A_{9,1}^2]$ & 55   \\
\makebox[0pt][l]{\fboxsep0pt\colorbox{Mygrey} {\strut\hspace*{1.12\linewidth}}}
14. & $8m+7$ & \tiny{$\left(\frac{1}{2},\frac{8m+7}{8}\right)$} & \tiny{128$m^2$ + 216$m$ + 91} & $D_{8m+7,1}$ & $17$ & \tiny{$\left(\frac{3}{2},\frac{9}{8}\right)$} & \tiny{$221$} & $\mathcal{E}_3[A_{11,1}E_{6,1}]$ & 59 \\
\makebox[0pt][l]{\fboxsep0pt\colorbox{Mywhite} {\strut\hspace*{1.12\linewidth}}}
15. & $8m+9$ & \tiny{$\left(\frac{1}{2},\frac{8m+9}{8}\right)$} & \tiny{128$m^2$ + 280$m$ + 153} & $D_{8m+9,1}$ & $15$ & \tiny{$\left(\frac{3}{2},\frac{7}{8}\right)$} & \tiny{$255$} & $\mathcal{E}_3[A_{15,1}]$ & 63 \\
\makebox[0pt][l]{\fboxsep0pt\colorbox{Mygrey} {\strut\hspace*{1.12\linewidth}}}
16. & $8m+10$ & \tiny{$\left(\frac{1}{2},\frac{8m+10}{8}\right)$} & \tiny{128$m^2$ + 312$m$ + 190} & $D_{8m+10,1}$ & $14$ & \tiny{$\left(\frac{3}{2},\frac{3}{4}\right)$} & \tiny{$266$} & $E_{7,1}^{\oplus 2}$ & 64 \\
\makebox[0pt][l]{\fboxsep0pt\colorbox{Mywhite} {\strut\hspace*{1.12\linewidth}}}
17. & $8m+12$ & \tiny{$\left(\frac{1}{2},\frac{8m+12}{8}\right)$} & \tiny{128$m^2$ + 376$m$ + 276} & $D_{8m+12,1}$ & $12$ & \tiny{$\left(\frac{3}{2},\frac{1}{2}\right)$} & \tiny{$276$} & $D_{12,1}$ & 66 \\
\hline
\hline
\end{longtable}

The new meromorphic theories predicted by the above coset relations are summarised in Table \ref{t4}. The first 15 rows have $B_{r,1}$ factors and the next 19 rows have $D_{r,1}$ factors. We provide the Kac-Moody algebra of which the meromorphic theory is an extension (in the first row, the extension is of a combination of a Kac-Moody algebra with the Baby Monster module). When the integer $m$ is equal to 0 the theory is part of the list in \cite{Schellekens:1992db} and we provide the serial number of that list where this entry can be found. For all $m\ge 1$ the theory is new, to our knowledge. 

Finally, for each case we provide the linear combination of character bilinears corresponding to the character of the extension. Here we have exhibited the way in which the 9 characters of the theory $\cC\oplus \tcC$ are combined into a meromorphic extension ${\cal E}_1[\cC\oplus \tcC]$ (in special cases where $\cC=\tcC$ we have 6 rather than 9 characters for $\cC\oplus\tcC$, but the idea is the same). The quantities $\hchi_0,\hchi_2,\hchi_{m+2}$, labelled by their conformal dimensions, are respectively the bilinears $\chi_0\tchi_0,\chi_1\tchi_1,\chi_2\tchi_2$ of the characters of $\cC,\tcC$ (these however are labelled serially as $0,1,2$ rather than by their conformal dimensions, the latter can be read off from the table) \footnote{Note that the $\tchi_i$ are not characters of the Kac-Moody algebra, but of its three-character extension.}. The $\hchi$'s are three of the 9 (or 6) characters in $\cC\oplus \tcC$. In general the bilinear identity involves coefficients $d_1,d_2$ (recall that $d_0=1$). Thus
the column specifies $\hchi_0+d_1\hchi_1+d_2\hchi_2$ which  defines the new meromorphic theory ${\cal E}_1[\cC\oplus \tcC]$.

\setlength\LTleft{-50pt}
\setlength\LTright{0pt}
\begin{longtable}{l|ccc||c|ccc}
\caption{The $34$ infinite series of new meromorphic CFTs. In each series, $m=0$ corresponds to a Schellekens theory, whose Schellekens number S\# is given in the second last column. BM denotes the Baby Monster CFT. Each theory has central charge $8m+24$.}
\label{t4}\\
\hline
\hline
\makebox[0pt][l]{\fboxsep0pt\colorbox{Mywhite} {\strut\hspace*{1.21\linewidth}}}
\# & $\mathcal{H}$ &  S\# & \tiny{Modular} & \# & $\mathcal{H}$ & S\# & \tiny{Modular} \\
\makebox[0pt][l]{\fboxsep0pt\colorbox{Mywhite} {\strut\hspace*{1.21\linewidth}}}
&  &  & \tiny{invariant}& &  &  & \tiny{invariant} \\
\hline
\makebox[0pt][l]{\fboxsep0pt\colorbox{Mywhite} {\strut\hspace*{1.21\linewidth}}}
1. & $\mathcal{E}_1[B_{8m,1} \text{BM}]$  &  $0$ & \tiny{$\hchi_0 + \hchi_2 + \, \hchi_{m+2}$} &  2. & $\mathcal{E}_1[B_{8m+1,1} A_{1,2}^{\oplus 15}]$ &  $5$ & \tiny{$\hchi_0 + \hchi_2 + \,1024 \hchi_{m+2}$} \\
\makebox[0pt][l]{\fboxsep0pt\colorbox{Mygrey} {\strut\hspace*{1.21\linewidth}}}
3. & $\mathcal{E}_1[B_{8m+1,1} A_{3,4}^{\oplus 3}]$  & $7$  &  \tiny{$\hchi_0 + \hchi_2 + \,1024 \hchi_{m+2}$} & 4. & $\mathcal{E}_1[B_{8m+1,1} A_{5,6} C_{2,3}]$ & $8$ & \tiny{$\hchi_0 + \hchi_2 + \,1024 \hchi_{m+2}$}  \\
\makebox[0pt][l]{\fboxsep0pt\colorbox{Mywhite} {\strut\hspace*{1.21\linewidth}}}
5. & $\mathcal{E}_1[B_{8m+1,1} D_{5,8}]$  & $10$  &   \tiny{$\hchi_0 + \hchi_2 + \,1024 \hchi_{m+2}$} & 6. & $\mathcal{E}_1[B_{8m+2,1}C_{2,1}^{\oplus 3}  D_{4,2}^{\oplus 2}]$ & $25$ &  \tiny{$\hchi_0 + \hchi_2 + \,512 \hchi_{m+2}$} \\
\makebox[0pt][l]{\fboxsep0pt\colorbox{Mygrey} {\strut\hspace*{1.21\linewidth}}}
7. & $\mathcal{E}_1[B_{8m+2,1} A_{2,1}^{\oplus 2} A_{5,2}^{\oplus 2}]$  & $26$  &   \tiny{$\hchi_0 + \hchi_2 + \,512 \hchi_{m+2}$} & 8. & $\mathcal{E}_1[B_{8m+2,1} A_{2,1} E_{6,4}]$ & $28$ &  \tiny{$\hchi_0 + \hchi_2 + \,512 \hchi_{m+2}$}\\
\makebox[0pt][l]{\fboxsep0pt\colorbox{Mywhite} {\strut\hspace*{1.21\linewidth}}}
9. & $\mathcal{E}_1[B_{8m+3,1} B_{3,1} C_{4,1} D_{6,2}]$  & $39$  &   \tiny{$\hchi_0 + \hchi_2 + \,256 \hchi_{m+2}$} & 10. & $\mathcal{E}_1[B_{8m+3,1} A_{4,1} A_{9,2}]$ & $40$ &   \tiny{$\hchi_0 + \hchi_2 + \,256 \hchi_{m+2}$} \\
\makebox[0pt][l]{\fboxsep0pt\colorbox{Mygrey} {\strut\hspace*{1.21\linewidth}}}
11. & $\mathcal{E}_1[B_{8m+4,1} B_{4,1} D_{8,2}]$  & $47$  &    \tiny{$\hchi_0 + \hchi_2 + \,128 \hchi_{m+2}$} & 12. & $\mathcal{E}_1[B_{8m+4,1}\,C_{6,1}^{\oplus 2}]$ & $48$ &  \tiny{$\hchi_0 + \hchi_2 + \,128 \hchi_{m+2}$} \\
\makebox[0pt][l]{\fboxsep0pt\colorbox{Mywhite} {\strut\hspace*{1.21\linewidth}}}
13. & $\mathcal{E}_1[B_{8m+5,1}\,E_{7,2}\,F_{4,1}]$  & $53$  &  \tiny{$\hchi_0 + \hchi_2 + \hchi_{m+2}$} & 14. & $\mathcal{E}_1[B_{8m+6,1}\,C_{10,1}]$ & $56$ & \tiny{$\hchi_0 + \hchi_2 + 32 \, \hchi_{m+2}$} \\
\makebox[0pt][l]{\fboxsep0pt\colorbox{Mygrey} {\strut\hspace*{1.21\linewidth}}}
15. & $\mathcal{E}_1[B_{8m+8,1}\,E_{8,2}]$  & $62$  &  \tiny{$\hchi_0 + \hchi_2 + \hchi_{m+2}$}  \\
\hline
\hline
\makebox[0pt][l]{\fboxsep0pt\colorbox{Mywhite} {\strut\hspace*{1.21\linewidth}}}
16. & $\mathcal{E}_1[D_{8m+1,1}D_{1,1}^{\oplus 23}]$  & $1$  & \tiny{$\hchi_0 + 8\hchi_2 + 4096 \, \hchi_{m+2}$}  & 17. & $\mathcal{E}_1[D_{8m+2,1}A_{1,1}^{\oplus 22}]$ & $15$ & \tiny{$\hchi_0 + 64\hchi_2 + 4096 \, \hchi_{m+2}$} \\
\makebox[0pt][l]{\fboxsep0pt\colorbox{Mygrey} {\strut\hspace*{1.21\linewidth}}}
18. & $\mathcal{E}_1[D_{8m+2,1}A_{1,1}^{\oplus 2}A_{3,2}^{\oplus 4}]$  & $16$  & \tiny{$\hchi_0 + 64\hchi_2 + 4096 \, \hchi_{m+2}$}  & 19. & $\mathcal{E}_1[D_{8m+2,1}A_{1,1}A_{5,3}D_{4,3}]$ & $17$ & \tiny{$\hchi_0 + 64\hchi_2 + 4096 \, \hchi_{m+2}$} \\
\makebox[0pt][l]{\fboxsep0pt\colorbox{Mywhite} {\strut\hspace*{1.21\linewidth}}}
20. & $\mathcal{E}_1[D_{8m+2,1}A_{1,1}A_{7,4}]$  & $18$  &  \tiny{$\hchi_0 + 64\hchi_2 + 4096 \, \hchi_{m+2}$} & 21. & $\mathcal{E}_1[D_{8m+2,1}C_{3,2}D_{5,4}]$ & $19$ & \tiny{$\hchi_0 + 64\hchi_2 + 4096 \, \hchi_{m+2}$} \\
\makebox[0pt][l]{\fboxsep0pt\colorbox{Mygrey} {\strut\hspace*{1.21\linewidth}}}
22. & $\mathcal{E}_1[D_{8m+2,1}D_{6,5}]$  & $20$  &  \tiny{$\hchi_0 + 64\hchi_2 + 4096 \, \hchi_{m+2}$} & 23. & $\mathcal{E}_1[D_{8m+3,1}A_{3,1}^{\oplus 7}]$ & $30$ & \tiny{$\hchi_0 + 8\hchi_2 + 1024 \, \hchi_{m+2}$} \\
\makebox[0pt][l]{\fboxsep0pt\colorbox{Mywhite} {\strut\hspace*{1.21\linewidth}}}
24. & $\mathcal{E}_1[D_{8m+3,1}A_{3,1}D_{5,2}^{\oplus 2}]$  & $31$  &  \tiny{$\hchi_0 + 8\hchi_2 + 1024 \, \hchi_{m+2}$} & 25. & $\mathcal{E}_1[D_{8m+3,1}A_{7,2}C_{3,1}^{\oplus 2}]$ & $33$ & \tiny{$\hchi_0 + 8\hchi_2 + 1024 \, \hchi_{m+2}$} \\
\makebox[0pt][l]{\fboxsep0pt\colorbox{Mygrey} {\strut\hspace*{1.21\linewidth}}}
26. & $\mathcal{E}_1[D_{8m+3,1}\,D_{7,3}G_{2,1}]$  & $34$  &  \tiny{$\hchi_0 + 8\hchi_2 + 1024 \, \hchi_{m+2}$} & 27. & $\mathcal{E}_1[D_{8m+3,1}C_{7,2}]$ & $35$ & \tiny{$\hchi_0 + 8\hchi_2 + 1024 \, \hchi_{m+2}$} \\
\makebox[0pt][l]{\fboxsep0pt\colorbox{Mywhite} {\strut\hspace*{1.21\linewidth}}}
28. & $\mathcal{E}_1[D_{8m+5,1}A_{7,1}^{\oplus 2}D_{5,1}]$  & $49$  &  \tiny{$\hchi_0 + 8\hchi_2 + 256 \, \hchi_{m+2}$} & 29. & $\mathcal{E}_1[D_{8m+6,1}D_{6,1}^{\oplus 3}]$ & $54$ & \tiny{$\hchi_0 + 64\hchi_2 + 256 \, \hchi_{m+2}$} \\
\makebox[0pt][l]{\fboxsep0pt\colorbox{Mygrey} {\strut\hspace*{1.21\linewidth}}}
30. & $\mathcal{E}_1[D_{8m+6,1}A_{9,1}^{\oplus 2}]$  & $55$  &  \tiny{$\hchi_0 + 64\hchi_2 + 256 \, \hchi_{m+2}$} & 31. & $\mathcal{E}_1[D_{8m+7,1}A_{11,1}E_{6,1}]$ & $59$ & \tiny{$\hchi_0 + 8\hchi_2 + 64 \, \hchi_{m+2}$} \\
\makebox[0pt][l]{\fboxsep0pt\colorbox{Mywhite} {\strut\hspace*{1.21\linewidth}}}
32. & $\mathcal{E}_1[D_{8m+9,1}A_{15,1}]$  & $63$  &  \tiny{$\hchi_0 + 8\hchi_2 + 16 \, \hchi_{m+2}$} & 33. & $\mathcal{E}_1[D_{8m+10,1}E_{7,1}^{\oplus 2}]$ & $64$ & \tiny{$\hchi_0 + \hchi_2 + 2 \, \hchi_{m+2}$} \\
\makebox[0pt][l]{\fboxsep0pt\colorbox{Mygrey} {\strut\hspace*{1.21\linewidth}}}
34. & $\mathcal{E}_1[D_{8m+12,1}D_{12,1}]$  & $66$  &  \tiny{$\hchi_0 + \hchi_2 + \hchi_{m+2}$}  \\
\hline
\hline
\end{longtable}

We now work out an example in detail. This is a typical case in this table and should clarify the procedure that has been used for all cases. We pick Row 6 of Table \ref{t1}, involving the coset pairing of $B_{8m+5,1}$ with ${\cal E}_3[E_{7,2}\oplus F_{4,1}]$. The resulting meromorphic theory is in row 13 of Table \ref{t4}. Notice that $B_{r,1}$ is a three-character affine theory for all $r$ (the same is true for the $D_{r,1}$ series, see Appendix A for details).
We now claim that the two factors pair to a meromorphic character with $c^{\mathcal{H}}=8(m+3)$ and $(n_1,n_2)=(2,m+2)$ and that the bilinear identity is:
\begin{align}
    \chi_0^{\mathcal{H}} & = \chi_0\tilde{\chi}_0 + \chi_{\frac{1}{2}}\tilde{\chi}_\frac{3}{2} + \chi_{\frac{16m+11}{16}}\tilde{\chi}_{\frac{21}{16}} = \hchi_0 + \hchi_2 + \hchi_{m+2} \label{b8m5_ghm185_9char}
\end{align}
The notation has been explained above. It should be kept in mind that for $m=0$ the last two terms in the last expression remain distinct although both would be denoted by $\hchi_2$, one comes from the bilinear in $h=\half,\tilh=\frac32$ while the other comes from the bilinear in $h=\frac{11}{16},\tilh=\frac{21}{16}$.\footnote{However for self-dual pairs the two $\hchi_2$'s are indeed same at $m=0$.} However for $m\ge 1$, the case of interest here, there is no ambiguity in the notation.

Now we explain how the bilinear identity is proved for all $m$, with the $m$-independent coefficients $(d_1,d_2)=(1,1)$ in this family.
To start with, we have explicitly verified the bilinear relation, to order $q^{2000}$ in the $q$-series, for $32 \le c^\cH\le 72$, which corresponds to $1\le m\le 6$ by comparing the series expansion on both sides. We find $(d_1,d_2)=(1,1)$ in all these cases, and the modular invariant on the RHS of the bilinear relation to be:
\be
\begin{split}
    &c^{\mathcal{H}}=24: \, \, \, \chi^{\cal H}(\tau) = j - 504,  \\
    &c^{\mathcal{H}}=32: \, \, \, \chi^{\cal H}(\tau) = j^{\frac{4}{3}} - 456 \, j^{\frac{1}{3}},  \\
    &c^{\mathcal{H}}=40: \, \, \, \chi^{\cal H}(\tau) = j^{\frac{5}{3}} - 152 \, j^{\frac{2}{3}},  \\
    &c^{\mathcal{H}}=48: \, \, \, \chi^{\cal H}(\tau) = j^{2} + 408 \, j - 129024,  \\
    &c^{\mathcal{H}}=56: \, \, \, \chi^{\cal H}(\tau) = j^{\frac{7}{3}} + 1224 \, j^{\frac{4}{3}} - 374784 \, j^{\frac{1}{3}},  \\
    &c^{\mathcal{H}}=64: \, \, \, \chi^{\cal H}(\tau) = j^{\frac{8}{3}} + 2296 \, j^{\frac{5}{3}} - 659456 \, j^{\frac{2}{3}},  \\
    &c^{\mathcal{H}}=72: \, \, \, \chi^{\cal H}(\tau) = j^{3} + 3624 \, j^{2} - 839680 \, j - 33030144. \label{meros}
\end{split}
\ee
This led us to conjecture that $(d_1,d_2)=(1,1)$ independent of $m$, and then immediately to a proof of the conjecture. 

For the proof we switch to the notation of Eq.(\ref{gen_j}) and find a formula for the coefficients in this example for all $m$. For this we first write:
\begin{align}
    \hchi_0 + d_1 \, \hchi_2 + d_2 \, \hchi_{m+2} = q^{-\frac{c^{\mathcal{H}}}{24}}\left(\text{an integral power series of q}\right) \label{b8m5_ghm185_9char00}
\end{align}
After cancelling the fractional power of $q$ -- if any -- from both sides of the above equation, we see that:
    \begin{align}
        a_r(m) = & \, \text{coefficient of} \ q^r \ \text{in LHS of Eq.(\ref{b8m5_ghm185_9char00})}\nonumber\\
        - &\ \text{coefficient of} \ q^r \ \text{in} \ \left(j^{\frac{m+3}{3}} + a_1 \, j^{\frac{m}{3}} + \ldots + a_{r-1}j^{\frac{m+3}{3}-(r-1)}\right). \label{a_nb8m5}
    \end{align}
For $r=1$ we can write more explicitly:
\begin{align}
    a_1(m) = & \ \text{coefficient of} \ q \ \text{in LHS of Eq.(\ref{b8m5_ghm185_9char00})} \ - \ \text{coefficient of} \ q \ \text{in} \ j^{\frac{m+3}{3}} \nonumber\\
    = & \ 128 \, m^2 + 168 \, m + 240 - 248(m+3). \label{a_1b8m5}
\end{align}
Note that though $a_1(m)$ has been derived from Eq.(\ref{b8m5_ghm185_9char00}), it only comes from comparing $\mathcal{O}(q)$ coefficient on both sides of Eq.(\ref{b8m5_ghm185_9char00}) and hence the formula for $a_1(m)$ is independent of $d_1$ and $d_2$. This is because on the LHS of Eq.(\ref{b8m5_ghm185_9char00}) $d_1$ appears at $\mathcal{O}(q^2)$ and $d_2$ appears at $\mathcal{O}(q^{m+2})$.

The characters $\chi_i$ of the $B$-series affine theory at level 1 are known to be given by Jacobi $\theta$-constants.
Using this representation one can show that $m_2(m)$ is quartic in $m$ (recall that $m_2$ is the second-level degeneracy for the identity character). This in turn would imply that the coefficient of $q^2$ in $\chi^\cH$ would be quartic in $m$ and hence $a_2$ would be quartic in $m$. Using this, we now prove that $d_1(m)$ is independent of $m$. 

We first employ {\it Mathematica} to solve (for $32\leq c^{\cal H}\leq 72$ as in Eq.(\ref{meros}) above):
\begin{align}
    (m,\text{coefficient of} \ q^2 \ \text{in} \ \chi^\cH) = &[(1,272432), (2,560268), (3,1121832), (4,2159876), \nonumber\\ 
    &(5,3942688), (6,6804092)]. \label{dataset_cq2}
\end{align}
which returns:
\begin{align}
    \text{coefficient of} \ q^2 \ \text{in } \chi^\cH = 121108 + \frac{337268}{3} \, m + \frac{89056}{3} \, m^2 + \frac{19456}{3} \, m^3 + \frac{8192}{3} \, m^4. \label{formula_cq2}
\end{align}
From the above we can get a general formula for $a_2$ using Eq.(\ref{gen_j}),
\begin{align}
    a_2(m) = &121108 + \frac{337268}{3} \, m + \frac{89056}{3} \, m^2 + \frac{19456}{3} \, m^3 + \frac{8192}{3} \, m^4 \nonumber\\
    & - 4124 (m+3) - 30752 (m+2) (m+3) - 248 \, m \, a_1. \label{formula_a2}
\end{align}
Comparing the $\mathcal{O}(q^2)$ term on both sides of  Eq.(\ref{bilinrel})), we find:
\begin{align}
    &\tilde{m}_2 + \tilde{m}_1 m_1(m) + m_2(m) + d_1(m) \, D_1(m)\tilde{D}_1 \nonumber\\
    &= \, a_2(m) + 248 \, m \, a_1(m) + 4124 (m+3) + 30752 (m+2) (m+3). \label{q2_coeff}
\end{align}    
Now from row 6 of Table \ref{t1} we read off:
\be
\begin{split}
&{\tilde m}_1 = 185\\
&m_1(m) = 128m^2+168m+55, \label{m1m}
\end{split}
\ee
while from the $\theta$-constant representation of the characters of $B_{r,1}$ we get:
\be
\begin{split}
D_1(r)&=2r+1=16m+11\\
m_2(r) &= 1+\frac{25}{6}r+\frac{23}{6}r^2-\frac23r^3+\frac23 r^4, \label{m2m}
\end{split}
\ee
where $r=8m+5$ for the present example.

Finally, solving the MLDE for the ($m$-independent) $\tilde {\cal C}$ theory in row 6 of Table \ref{t1} gives:
\be
\begin{split}
{\tilde m_2} &=56351,\qquad {\tilde D}_1 = 4921. \label{tilm2}
\end{split}
\ee
Inserting Eqs.(\ref{m1m}), (\ref{m2m}), (\ref{tilm2}) in \eref{q2_coeff}, we get:
\begin{align}
    56351 + &185(128 \, m^2 + 168 \, m + 55)
    + \left(1 + \frac{25}{6}(8m+5) + \frac{23}{6}(8m+5)^2 \right. \nonumber\\
    &\left.-\frac{2}{3}(8m+5)^3 + \frac{2}{3}(8m+5)^4\right) + d_1(m)4921(16m+11) \nonumber\\
    =& \, a_2(m) + 248 \, m \, a_1(m) + 4124 (m+3) + 30752 (m+2) (m+3)
\end{align}    
from which, using \eref{formula_a2} and \eref{a_1b8m5}, we find that $d_1(m)=1$ for all $m$. Similarly one can argue that $d_2=1$ $\forall \, m\geq 1$.

The corresponding results for the remaining infinite families can be explained in the same way. In all cases the coefficients $d_1,d_2$ in the bilinear relation for all $m$ are the same as those obtained for $m=0$, i.e. the $c^\cH=24$ case. Thus we have verified (still to order $q^{2000}$ in the $q$-series), that this new theory satisfies a bilinear relation with $\tcC$ forming a modular invariant with $c^\cH=8(m+3)$ for all $m$. 

Now let us give below, explicitly, the $3$-character extension of the twelve character theory $E_{7,2}F_{4,1}$,
\begin{align}
    &\tilde{\chi}_0 = \chi^E_0\chi^F_0 + \chi^E_{\frac{7}{5}}\chi^F_{\frac{3}{5}} \nonumber\\
    &\tilde{\chi}_{\frac{3}{2}} = \chi^E_{\frac{3}{2}}\chi^F_0 + \chi^E_{\frac{9}{10}}\chi^F_{\frac{3}{5}} \nonumber\\
    &\tilde{\chi}_{\frac{21}{16}} = \chi^E_{\frac{21}{16}}\chi^F_0 + \chi^E_{\frac{57}{80}}\chi^F_{\frac{3}{5}}
\end{align}
where $\chi^E_i$'s represent the six characters of $E_{7,2}$ and $\chi^F_i$s represent the two characters of $F_{4,1}$. The above expressions are obtained by comparing the characters of the extension (found from the MLDE approach) with the leading behaviour of the characters $\chi^E,\chi^F$ which is given by the dimensions of integrable representations.

Using the above result, now we can explicitly write the $1$-character extension of the $36$-character theory $B_{8m+5,1}E_{7,2}F_{4,1}$:
\begin{align}
    \chi^{\mathcal{H}} &= \hchi_0 + \hchi_{2} + \hchi_{m+2} = \chi_0\tilde{\chi}_0 + \chi_\frac{1}{2}\tilde{\chi}_{\frac{3}{2}} + \chi_{\frac{16m+11}{16}}\tilde{\chi}_{\frac{21}{16}}\nonumber \\
    &= j^{1/3}\left(j+128 \, m^2+168 \, m+240 - 248(m+3)\right),  \nonumber \\
    &= \chi_0\chi^E_0\chi^F_0 + \chi_0\chi^E_{\frac{7}{5}}\chi^F_{\frac{3}{5}} + \chi_{\frac{1}{2}}\chi^E_{\frac{3}{2}}\chi^F_0 + \chi_{\frac{1}{2}}\chi^E_{\frac{9}{10}}\chi^F_{\frac{3}{5}} + \chi_{\frac{16m+11}{16}}\chi^E_{\frac{21}{16}}\chi^F_0 + \chi_{\frac{16m+11}{16}}\chi^E_{\frac{57}{80}}\chi^F_{\frac{3}{5}}, \label{b13_ghm185_36char11789}
\end{align}
where $\hchi_i$s represent the nine characters of $B_{8m+5,1}\oplus\mathcal{E}_{3}[E_{7,2}F_{4,1}]$, $\chi_i$'s represent the three characters of $B_{8m+5,1}$, $\tilde{\chi}_i$'s represent the three characters of $\mathcal{E}_3[E_{7,2}F_{4,1}]$.

\subsection{New meromorphic theories at $c=32,40$\,: finite families}

In this section we exhibit a finite set of novel meromorphic theories with central charges $c^{\mathcal{H}}=32$ and $c^{\mathcal{H}}=40$ only. As in the previous cases, the bilinear relations for these examples have also been verified to order $q^{2000}$. Tables \ref{t2} and \ref{t3} exhibit the coset pairings which correspond respectively to $c^{\mathcal{H}}=32$ with $(n_1,n_2)=(2,3)$, and $c^{\mathcal{H}}=40$ with $(n_1,n_2)=(3,3)$. There exists no new family for $c^{\mathcal{H}}=40$ with $(n_1,n_2)=(2,4)$. At $c^{\mathcal{H}}=32$ we find that there are 7 additional meromorphic theories. Out of these, 3 are novel non-lattice meromorphic theories (second and third line of row 1, and row 2). The remaining 4 are lattice meromorphic theories whose lattices can be found in \cite{2003MaCom..72..839K}. Similarly, at $c^{\mathcal{H}}=40$, there are 39 additional meromorphic theories, of which 32 are novel non-lattice theories and the remaining 7 are lattice theories.

\setlength\LTleft{60pt}
\setlength\LTright{0pt}
\begin{longtable}{l||cccc||cccc}
\caption{Coset relations for $c^\mathcal{H}=32$ with $(n_1,n_2)=(2,3)$.}
\label{t2}\\
\hline
\hline
\makebox[0pt][l]{\fboxsep0pt\colorbox{Mywhite} {\strut\hspace*{0.71\linewidth}}}
\# & $c$ & \tiny{$(h_1,h_2)$} & \tiny{$m_1$} & $\mathcal{C}$ & $\Tilde{c}$ & \tiny{$(\Tilde{h}_1,\Tilde{h}_2)$} & \tiny{$\Tilde{m}_1$} & $\Tilde{\mathcal{C}}$ \\
\hline
\makebox[0pt][l]{\fboxsep0pt\colorbox{Mywhite} {\strut\hspace*{0.71\linewidth}}}
1. & $12$ & \tiny{$\left(\frac{2}{3},\frac{4}{3}\right)$} & \tiny{156} & $E_{6,1}^{\oplus 2}$ & $20$ & \tiny{$\left(\frac{4}{3},\frac{5}{3}\right)$} & \tiny{$80$} & $\mathcal{E}_3[A_{2,1}^{\oplus 10}]$ \\
\makebox[0pt][l]{\fboxsep0pt\colorbox{Mywhite} {\strut\hspace*{0.71\linewidth}}}
  &  &  &  &  &  &  &  & $\mathcal{E}_3[A_{5,2}^{\oplus 2}C_{2,1}]$ \\
\makebox[0pt][l]{\fboxsep0pt\colorbox{Mywhite} {\strut\hspace*{\linewidth}}}
  &  &  &  &  &  &  &  & $\mathcal{E}_3[A_{8,3}]$ \\  
\makebox[0pt][l]{\fboxsep0pt\colorbox{Mygrey} {\strut\hspace*{0.71\linewidth}}}
2. & $\frac{68}{5}$ & \tiny{$\left(\frac{4}{5},\frac{7}{5}\right)$} & \tiny{136} & $\mathcal{E}_3[C_{8,1}]$ & $\frac{92}{5}$ & \tiny{$\left(\frac{6}{5},\frac{8}{5}\right)$} & \tiny{$92$} & $\mathcal{E}_3[E_{6,3}G_{2,1}]$ \\
\makebox[0pt][l]{\fboxsep0pt\colorbox{Mywhite} {\strut\hspace*{0.71\linewidth}}}
3. & $14$ & \tiny{$\left(\frac{3}{4},\frac{3}{2}\right)$} & \tiny{266} & $E_{7,1}^{\oplus 2}$ & $18$ & \tiny{$\left(\frac{5}{4},\frac{3}{2}\right)$} & \tiny{$198$} & $\mathcal{E}_3[D_{6,1}^{\oplus 3}]$ \\
\makebox[0pt][l]{\fboxsep0pt\colorbox{Mywhite} {\strut\hspace*{0.71\linewidth}}}
  &  &  &  &  &  &  &  & $\mathcal{E}_3[A_{9,1}^{\oplus 2}]$ \\
\makebox[0pt][l]{\fboxsep0pt\colorbox{Mygrey} {\strut\hspace*{0.71\linewidth}}}
4. & $15$ & \tiny{$\left(\frac{7}{8},\frac{3}{2}\right)$} & \tiny{255} & $\mathcal{E}_3[A_{15,1}]$ & $17$ & \tiny{$\left(\frac{9}{8},\frac{3}{2}\right)$} & \tiny{$221$} & $\mathcal{E}_3[A_{11,1}E_{6,1}]$ \\
\hline
\hline
\end{longtable}

\newpage

\setlength\LTleft{40pt}
\setlength\LTright{0pt}
\begin{longtable}{l||cccc||cccc}
\caption{Coset relations for $c^\mathcal{H}=40$ with $(n_1,n_2)=(3,3)$.}
\label{t3}\\
\hline
\hline
\makebox[0pt][l]{\fboxsep0pt\colorbox{Mywhite} {\strut\hspace*{0.79\linewidth}}}
\# & $c$ & \tiny{$(h_1,h_2)$} & \tiny{$m_1$} & $\mathcal{C}$ & $\Tilde{c}$ & \tiny{$(\Tilde{h}_1,\Tilde{h}_2)$} & \tiny{$\Tilde{m}_1$} & $\Tilde{\mathcal{C}}$ \\
\makebox[0pt][l]{\fboxsep0pt\colorbox{Mywhite} {\strut\hspace*{0.79\linewidth}}}
  &  &  &  &  &  &  &  &  \\
\hline
\hline
\makebox[0pt][l]{\fboxsep0pt\colorbox{Mywhite} {\strut\hspace*{0.79\linewidth}}}
1. & $20$ & \tiny{$\left(\frac{1}{2},\frac{5}{2}\right)$} & \tiny{780} & $D_{20,1}$ & $20$ & \tiny{$\left(\frac{5}{2},\frac{1}{2}\right)$} & \tiny{$780$} & $D_{20,1}$ \\
\makebox[0pt][l]{\fboxsep0pt\colorbox{Mygrey} {\strut\hspace*{0.79\linewidth}}}
2. & $20$ & \tiny{$\left(\frac{4}{3},\frac{5}{3}\right)$} & \tiny{80} & $\mathcal{E}_3[A_{2,1}^{\oplus 10}]$ & $20$ & \tiny{$\left(\frac{5}{3},\frac{4}{3}\right)$} & \tiny{$80$} & $\mathcal{E}_3[A_{2,1}^{\oplus 10}]$ \\
\makebox[0pt][l]{\fboxsep0pt\colorbox{Mygrey} {\strut\hspace*{0.79\linewidth}}}
 &  &  &  & $\mathcal{E}_3[A_{2,1}^{\oplus 10}]$ &  &  &  & $\mathcal{E}_3[A_{5,2}^{\oplus 2}C_{2,1}]$ \\
\makebox[0pt][l]{\fboxsep0pt\colorbox{Mygrey} {\strut\hspace*{0.79\linewidth}}}
  &  &  &  & $\mathcal{E}_3[A_{2,1}^{\oplus 10}]$ &  &  &  & $\mathcal{E}_3[A_{8,3}]$ \\ 
\makebox[0pt][l]{\fboxsep0pt\colorbox{Mygrey} {\strut\hspace*{0.79\linewidth}}}
  &  &  &  & $\mathcal{E}_3[A_{5,2}^{\oplus 2}C_{2,1}]$ &  &  &  & $\mathcal{E}_3[A_{5,2}^{\oplus 2}C_{2,1}]$ \\ 
\makebox[0pt][l]{\fboxsep0pt\colorbox{Mygrey} {\strut\hspace*{0.79\linewidth}}}
  &  &  &  & $\mathcal{E}_3[A_{5,2}^{\oplus 2}C_{2,1}]$ &  &  &  & $\mathcal{E}_3[A_{8,3}]$ \\
\makebox[0pt][l]{\fboxsep0pt\colorbox{Mygrey} {\strut\hspace*{0.79\linewidth}}}
  &  &  &  & $\mathcal{E}_3[A_{8,3}]$ &  &  &  & $\mathcal{E}_3[A_{8,3}]$ \\
\makebox[0pt][l]{\fboxsep0pt\colorbox{Mywhite} {\strut\hspace*{0.79\linewidth}}}
3. & $20$ & \tiny{$\left(\frac{7}{5},\frac{8}{5}\right)$} & \tiny{120} & $\mathcal{E}_3[A_{4,1}^{\oplus 5}]$ & $20$ & \tiny{$\left(\frac{8}{5},\frac{7}{5}\right)$} & \tiny{$120$} & $\mathcal{E}_3[A_{4,1}^{\oplus 5}]$ \\
\makebox[0pt][l]{\fboxsep0pt\colorbox{Mywhite} {\strut\hspace*{0.79\linewidth}}}
  &  &  &  & $\mathcal{E}_3[A_{4,1}^{\oplus 5}]$ &  &  &  & $\mathcal{E}_3[A_{9,2}B_{3,1}]$\\
\makebox[0pt][l]{\fboxsep0pt\colorbox{Mywhite} {\strut\hspace*{0.79\linewidth}}}
  &  &  &  & $\mathcal{E}_3[A_{9,2}B_{3,1}]$ &  &  &  & $\mathcal{E}_3[A_{9,2}B_{3,1}]$\\  
\makebox[0pt][l]{\fboxsep0pt\colorbox{Mygrey} {\strut\hspace*{0.79\linewidth}}}
4. & $17$ & \tiny{$\left(\frac{3}{2},\frac{9}{8}\right)$} & \tiny{221} & $\mathcal{E}_3[A_{11,1}E_{6,1}]$ & $23$ & \tiny{$\left(\frac{3}{2},\frac{15}{8}\right)$} & \tiny{$23$} & $\mathcal{E}_3[D_{1,1}^{\oplus 23}]$ \\
\makebox[0pt][l]{\fboxsep0pt\colorbox{Mywhite} {\strut\hspace*{0.79\linewidth}}}
5. & $\frac{35}{2}$ & \tiny{$\left(\frac{3}{2},\frac{19}{16}\right)$} & \tiny{210} & $\mathcal{E}_3[C_{10,1}]$ & $\frac{45}{2}$ & \tiny{$\left(\frac{3}{2},\frac{29}{16}\right)$} & \tiny{$45$} & $\mathcal{E}_3[A_{1,2}^{\oplus 15}]$ \\
\makebox[0pt][l]{\fboxsep0pt\colorbox{Mywhite} {\strut\hspace*{0.79\linewidth}}}
 &  &  &  & $\mathcal{E}_3[C_{10,1}]$ &  &  &  & $\mathcal{E}_3[A_{3,4}^{\oplus 3}]$ \\
\makebox[0pt][l]{\fboxsep0pt\colorbox{Mywhite} {\strut\hspace*{0.79\linewidth}}}
  &  &  &  & $\mathcal{E}_3[C_{10,1}]$ &  &  &  & $\mathcal{E}_3[A_{5,6}C_{2,3}]$ \\ 
\makebox[0pt][l]{\fboxsep0pt\colorbox{Mywhite} {\strut\hspace*{0.79\linewidth}}}
  &  &  &  & $\mathcal{E}_3[C_{10,1}]$ &  &  &  & $\mathcal{E}_3[D_{5,8}]$\\ 
\makebox[0pt][l]{\fboxsep0pt\colorbox{Mygrey} {\strut\hspace*{0.79\linewidth}}}  
6. & $18$ & \tiny{$\left(\frac{5}{4},\frac{3}{2}\right)$} & \tiny{198} & $\mathcal{E}_3[D_{6,1}^{\oplus 3}]$ & $22$ & \tiny{$\left(\frac{7}{4},\frac{3}{2}\right)$} & \tiny{$66$} & $\mathcal{E}_3[A_{1,1}^{\oplus 22}]$\\
\makebox[0pt][l]{\fboxsep0pt\colorbox{Mygrey} {\strut\hspace*{0.79\linewidth}}}
 &  &  &  & $\mathcal{E}_3[D_{6,1}^{\oplus 3}]$ &  &  &  & $\mathcal{E}_3[A_{3,2}^{\oplus 4} A_{1,1}^{\oplus 2}]$ \\
\makebox[0pt][l]{\fboxsep0pt\colorbox{Mygrey} {\strut\hspace*{0.79\linewidth}}}
  &  &  &  & $\mathcal{E}_3[D_{6,1}^{\oplus 3}]$ &  &  &  & $\mathcal{E}_3[A_{5,3}D_{4,3}A_{1,1}]$ \\
\makebox[0pt][l]{\fboxsep0pt\colorbox{Mygrey} {\strut\hspace*{0.79\linewidth}}}
  &  &  &  & $\mathcal{E}_3[D_{6,1}^{\oplus 3}]$ &  &  &  & $\mathcal{E}_3[A_{7,4}A_{1,1}]$ \\
\makebox[0pt][l]{\fboxsep0pt\colorbox{Mygrey} {\strut\hspace*{0.79\linewidth}}}
  &  &  &  & $\mathcal{E}_3[D_{6,1}^{\oplus 3}]$ &  &  &  & $\mathcal{E}_3[D_{5,4}C_{3,2}]$ \\  
\makebox[0pt][l]{\fboxsep0pt\colorbox{Mygrey} {\strut\hspace*{0.79\linewidth}}}
  &  &  &  & $\mathcal{E}_3[D_{6,1}^{\oplus 3}]$ &  &  &  & $\mathcal{E}_3[D_{6,5}]$ \\ 
\makebox[0pt][l]{\fboxsep0pt\colorbox{Mygrey} {\strut\hspace*{0.79\linewidth}}}
  &  &  &  & $\mathcal{E}_3[A_{9,1}^{\oplus 2}]$ &  &  &  & $\mathcal{E}_3[A_{1,1}^{\oplus 22}]$ \\
\makebox[0pt][l]{\fboxsep0pt\colorbox{Mygrey} {\strut\hspace*{0.79\linewidth}}}
  &  &  &  & $\mathcal{E}_3[A_{9,1}^{\oplus 2}]$ &  &  &  & $\mathcal{E}_3[A_{3,2}^{\oplus 4} A_{1,1}^{\oplus 2}]$ \\ 
\makebox[0pt][l]{\fboxsep0pt\colorbox{Mygrey} {\strut\hspace*{0.79\linewidth}}}
  &  &  &  & $\mathcal{E}_3[A_{9,1}^{\oplus 2}]$ &  &  &  & $\mathcal{E}_3[A_{5,3}D_{4,3}A_{1,1}]$ \\
\makebox[0pt][l]{\fboxsep0pt\colorbox{Mygrey} {\strut\hspace*{0.79\linewidth}}}
  &  &  &  & $\mathcal{E}_3[A_{9,1}^{\oplus 2}]$ &  &  &  & $\mathcal{E}_3[A_{7,4}A_{1,1}]$ \\
\makebox[0pt][l]{\fboxsep0pt\colorbox{Mygrey} {\strut\hspace*{0.79\linewidth}}}
  &  &  &  & $\mathcal{E}_3[A_{9,1}^{\oplus 2}]$ &  &  &  & $\mathcal{E}_3[D_{5,4}C_{3,2}]$ \\
\makebox[0pt][l]{\fboxsep0pt\colorbox{Mygrey} {\strut\hspace*{0.79\linewidth}}}
  &  &  &  & $\mathcal{E}_3[A_{9,1}^{\oplus 2}]$ &  &  &  & $\mathcal{E}_3[D_{6,5}]$ \\  
\makebox[0pt][l]{\fboxsep0pt\colorbox{Mywhite} {\strut\hspace*{0.79\linewidth}}}
7. & $\frac{37}{2}$ & \tiny{$\left(\frac{3}{2},\frac{21}{16}\right)$} & \tiny{185} & $\mathcal{E}_3[E_{7,2}F_{4,1}]$ & $\frac{43}{2}$ & \tiny{$\left(\frac{3}{2},\frac{27}{16}\right)$} & \tiny{$86$} & $\mathcal{E}_3[D_{4,2}^{\oplus 2}C_{2,1}^{\oplus 3}]$ \\
\makebox[0pt][l]{\fboxsep0pt\colorbox{Mywhite} {\strut\hspace*{0.79\linewidth}}}
 &  &  &  & $\mathcal{E}_3[E_{7,2}F_{4,1}]$ &  &  &  & $\mathcal{E}_3[A_{5,2}^{\oplus 2}A_{2,1}^{\oplus 2}]$\\
\makebox[0pt][l]{\fboxsep0pt\colorbox{Mywhite} {\strut\hspace*{0.79\linewidth}}}
  &  &  &  & $\mathcal{E}_3[E_{7,2}F_{4,1}]$ &  &  &  & $\mathcal{E}_3[E_{6,4}A_{2,1}]$ \\ 
\makebox[0pt][l]{\fboxsep0pt\colorbox{Mygrey} {\strut\hspace*{0.79\linewidth}}}  
8. & $19$ & \tiny{$\left(\frac{3}{2},\frac{11}{8}\right)$} & \tiny{171} & $\mathcal{E}_3[D_{5,1}A_{7,1}^{\oplus 2}]$ & $21$ & \tiny{$\left(\frac{3}{2},\frac{13}{8}\right)$} & \tiny{$105$} & $\mathcal{E}_3[A_{3,1}^{\oplus 7}]$ \\
\makebox[0pt][l]{\fboxsep0pt\colorbox{Mygrey} {\strut\hspace*{0.79\linewidth}}}
 &  &  &  & $\mathcal{E}_3[D_{5,1}A_{7,1}^{\oplus 2}]$ &  &  &  & $\mathcal{E}_3[D_{5,2}^{\oplus 2}A_{3,1}]$ \\
\makebox[0pt][l]{\fboxsep0pt\colorbox{Mygrey} {\strut\hspace*{0.79\linewidth}}}
  &  &  &  & $\mathcal{E}_3[D_{5,1}A_{7,1}^{\oplus 2}]$ &  &  &  & $\mathcal{E}_3[A_{7,2}C_{3,1}^{\oplus 2}]$ \\
\makebox[0pt][l]{\fboxsep0pt\colorbox{Mygrey} {\strut\hspace*{0.79\linewidth}}}
  &  &  &  & $\mathcal{E}_3[D_{5,1}A_{7,1}^{\oplus 2}]$ &  &  &  & $\mathcal{E}_3[D_{7,3}G_{2,1}]$ \\
\makebox[0pt][l]{\fboxsep0pt\colorbox{Mygrey} {\strut\hspace*{0.79\linewidth}}}
  &  &  &  & $\mathcal{E}_3[D_{5,1}A_{7,1}^{\oplus 2}]$ &  &  &  & $\mathcal{E}_3[C_{7,2}]$ \\
\makebox[0pt][l]{\fboxsep0pt\colorbox{Mywhite} {\strut\hspace*{0.79\linewidth}}}  
9. & $\frac{39}{2}$ & \tiny{$\left(\frac{3}{2},\frac{23}{16}\right)$} & \tiny{156} & $\mathcal{E}_3[D_{8,2}B_{4,1}]$ & $\frac{41}{2}$ & \tiny{$\left(\frac{3}{2},\frac{25}{16}\right)$} & \tiny{$123$} & $\mathcal{E}_3[D_{6,2}C_{4,1}B_{3,1}]$ \\
\makebox[0pt][l]{\fboxsep0pt\colorbox{Mywhite} {\strut\hspace*{0.79\linewidth}}}
 &  &  &  & $\mathcal{E}_3[D_{8,2}B_{4,1}]$ &  &  &  & $\mathcal{E}_3[A_{9,2}A_{4,1}]$ \\
\makebox[0pt][l]{\fboxsep0pt\colorbox{Mywhite} {\strut\hspace*{0.79\linewidth}}}
  &  &  &  & $\mathcal{E}_3[C_{6,1}^{\oplus 2}]$ &  &  &  & $\mathcal{E}_3[D_{6,2}C_{4,1}B_{3,1}]$ \\ 
\makebox[0pt][l]{\fboxsep0pt\colorbox{Mywhite} {\strut\hspace*{0.79\linewidth}}}
  &  &  &  & $\mathcal{E}_3[C_{6,1}^{\oplus 2}]$ &  &  &  & $\mathcal{E}_3[A_{9,2}A_{4,1}]$ \\ 
\hline
\hline
\end{longtable}

The new meromorphic theories predicted by the above coset relations are summarised in Table \ref{t5}. The format for the columns of this table is similar to that of Table \ref{t4}. The first 7 entries are theories at $c=32$ and the next 39 are theories at $c=40$.

\setlength\LTleft{-37pt}
\setlength\LTright{0pt}
\begin{longtable}{l|ccc||c|ccc}
\caption{The finite set of 46 new meromorphic CFTs at central charges $32$ and $40$.}
\label{t5}\\
\hline
\hline
\makebox[0pt][l]{\fboxsep0pt\colorbox{Mywhite} {\strut\hspace*{1.16\linewidth}}}
\# & $\mathcal{H}$ & $c^\mathcal{H}$ &  $\chi^{\mathcal{H}}$ & \# & $\mathcal{H}$ & $c^\mathcal{H}$ &  $\chi^{\mathcal{H}}$ \\
\hline
\hline
\makebox[0pt][l]{\fboxsep0pt\colorbox{Mywhite} {\strut\hspace*{1.16\linewidth}}}
1. &  $\mathcal{E}_1[A_{2,1}^{\oplus 10}E_{6,1}^{\oplus 2}]$ & $32$  &  \tiny{$\hchi_0 + 972\hchi_2 + 2^2\cdot 3^7\hchi_3$} &
2. &    $\mathcal{E}_1[A_{5,2}^{\oplus 2}C_{2,1}E_{6,1}^{\oplus 2}]$ & $32$  & \tiny{$\hchi_0 + 972\hchi_2 + 2^2\cdot 3^7\hchi_3$} \\
\makebox[0pt][l]{\fboxsep0pt\colorbox{Mygrey} {\strut\hspace*{1.16\linewidth}}}
3. &  $\mathcal{E}_1[A_{8,3}E_{6,1}^{\oplus 2}]$ & $32$  &  \tiny{$\hchi_0 + 972\hchi_2 + 2^2\cdot 3^7\hchi_3$} & 4. &   $\mathcal{E}_1[C_{8,1}E_{6,3}G_{2,1}]$ & $32$  & \tiny{$\hchi_0 + \hchi_2 + 1250\hchi_3$} \\
\makebox[0pt][l]{\fboxsep0pt\colorbox{Mywhite} {\strut\hspace*{1.16\linewidth}}}
5. & $\mathcal{E}_1[D_{6,1}^{\oplus 3}E_{7,1}^{\oplus 2}]$  & $32$  &  \tiny{$\hchi_0 + 256\hchi_2 + 64\hchi_3$} & 
6. &   $\mathcal{E}_1[A_{9,1}^{\oplus 2}E_{7,1}^{\oplus 2}]$ & $32$  & \tiny{$\hchi_0 + 256\hchi_2 + 64\hchi_3$} \\
\makebox[0pt][l]{\fboxsep0pt\colorbox{Mygrey} {\strut\hspace*{1.16\linewidth}}}
7. & $\mathcal{E}_1[A_{11,1}A_{15,1}E_{6,1}]$  & $32$  &  \tiny{$\hchi_0 + 512\hchi_2 + 64\hchi_3$} \\
\hline
\hline
\makebox[0pt][l]{\fboxsep0pt\colorbox{Mywhite} {\strut\hspace*{1.16\linewidth}}}
8. &   $\mathcal{E}_1[D_{20,1}^{\oplus 2}]$ & $40$  & \tiny{$\hchi_0 + 2\hchi_3$} & 
9. & $\mathcal{E}_1[A_{2,1}^{\oplus 20}]$  & $40$  &  \tiny{$\hchi_0 + 2^3\cdot 3^{12}\hchi_3$} \\
\makebox[0pt][l]{\fboxsep0pt\colorbox{Mygrey} {\strut\hspace*{1.16\linewidth}}}
10. & $\mathcal{E}_1[A_{2,1}^{\oplus 10}A_{5,2}^{\oplus 2}C_{2,1}]$  & $40$  &  \tiny{$\hchi_0 + 2^3\cdot 3^{12}\hchi_3$} & 
11. & $\mathcal{E}_1[A_{2,1}^{\oplus 10}A_{8,3}]$  & $40$  &  \tiny{$\hchi_0 + 2^3\cdot 3^{12}\hchi_3$} \\
\makebox[0pt][l]{\fboxsep0pt\colorbox{Mywhite} {\strut\hspace*{1.16\linewidth}}}
12. & $\mathcal{E}_1[A_{5,2}^{\oplus 4}C_{2,1}^{\oplus 2}]$  & $40$  &  \tiny{$\hchi_0 + 2^3\cdot 3^{12}\hchi_3$} & 
13. & $\mathcal{E}_1[A_{5,2}^{\oplus 2}A_{8,3}C_{2,1}]$  & $40$  &  \tiny{$\hchi_0 + 2^3\cdot 3^{12}\hchi_3$} \\
\makebox[0pt][l]{\fboxsep0pt\colorbox{Mygrey} {\strut\hspace*{1.16\linewidth}}}
14. & $\mathcal{E}_1[A_{8,3}^{\oplus 2}]$  & $40$  &  \tiny{$\hchi_0 + 2^3\cdot 3^{12}\hchi_3$} & 
15. & $\mathcal{E}_1[A_{4,1}^{\oplus 10}]$  & $40$  &  \tiny{$\hchi_0 + 2^2\cdot 5^{8}\hchi_3$} \\
\makebox[0pt][l]{\fboxsep0pt\colorbox{Mywhite} {\strut\hspace*{1.16\linewidth}}}
16. & $\mathcal{E}_1[A_{9,2}^{\oplus 2}B_{3,1}^{\oplus 2}]$  & $40$  &  \tiny{$\hchi_0 + 2^2\cdot 5^{8}\hchi_3$} & 
17. & $\mathcal{E}_1[A_{4,1}^{\oplus 5}A_{9,2}\,B_{3,1}]$  & $40$  &  \tiny{$\hchi_0 + 2^2\cdot 5^{8}\hchi_3$} \\
\makebox[0pt][l]{\fboxsep0pt\colorbox{Mygrey} {\strut\hspace*{1.16\linewidth}}}
18. & $\mathcal{E}_1[A_{11,1} D_{1,1}^{\oplus 23} E_{6,1}]$  & $40$  &  \tiny{$\hchi_0 + 2^6\hchi_3 + 2^{17}\hchi_3$} & 
19. & $\mathcal{E}_1[A_{1,2}^{\oplus 15} C_{10,1}]$  & $40$  &  \tiny{$\hchi_0 + \hchi_3 + 2^{15}\hchi_3$} \\
\makebox[0pt][l]{\fboxsep0pt\colorbox{Mywhite} {\strut\hspace*{1.16\linewidth}}}
20. & $\mathcal{E}_1[A_{3,4}^{\oplus 3}C_{10,1}]$  & $40$  &  \tiny{$\hchi_0 + \hchi_3 + 2^{15}\hchi_3$} &
21. & $\mathcal{E}_1[A_{5,6}C_{2,3}C_{10,1}]$  & $40$  &  \tiny{$\hchi_0 + \hchi_3 + 2^{15}\hchi_3$} \\
\makebox[0pt][l]{\fboxsep0pt\colorbox{Mygrey} {\strut\hspace*{1.16\linewidth}}}
22. & $\mathcal{E}_1[C_{10,1} D_{5,8}]$  & $40$  &  \tiny{$\hchi_0 + \hchi_3 + 2^{15}\hchi_3$} & 
23. & $\mathcal{E}_1[A_{1,1}^{\oplus 22}D_{6,1}^{\oplus 3}]$  & $40$  &  \tiny{$\hchi_0 + 2^{19}\hchi_3 + 2^{12}\hchi_3$} \\
\makebox[0pt][l]{\fboxsep0pt\colorbox{Mywhite} {\strut\hspace*{1.16\linewidth}}}
24. & $\mathcal{E}_1[A_{1,1}^{\oplus 2}A_{3,2}^{\oplus 4}D_{6,1}^{\oplus 3}]$  & $40$  &  \tiny{$\hchi_0 + 2^{19}\hchi_3 + 2^{12}\hchi_3$} &
25. & $\mathcal{E}_1[A_{1,1}A_{5,3}D_{4,3}D_{6,1}^{\oplus 3}]$  & $40$  &  \tiny{$\hchi_0 + 2^{19}\hchi_3 + 2^{12}\hchi_3$} \\
\makebox[0pt][l]{\fboxsep0pt\colorbox{Mygrey} {\strut\hspace*{1.16\linewidth}}}
26. & $\mathcal{E}_1[A_{1,1}A_{7,4}D_{6,1}^{\oplus 3}]$  & $40$  &  \tiny{$\hchi_0 + 2^{19}\hchi_3 + 2^{12}\hchi_3$} & 
27. & $\mathcal{E}_1[C_{3,2}D_{5,4}D_{6,1}^{\oplus 3}]$  & $40$  &  \tiny{$\hchi_0 + 2^{19}\hchi_3 + 2^{12}\hchi_3$} \\
\makebox[0pt][l]{\fboxsep0pt\colorbox{Mywhite} {\strut\hspace*{1.16\linewidth}}}
28. & $\mathcal{E}_1[D_{6,1}^{\oplus 3}D_{6,5}]$  & $40$  &  \tiny{$\hchi_0 + 2^{19}\hchi_3 + 2^{12}\hchi_3$} &
29. & $\mathcal{E}_1[A_{1,1}^{\oplus 22}A_{9,1}^{\oplus 2}]$  & $40$  &  \tiny{$\hchi_0 + 2^{19}\hchi_3 + 2^{12}\hchi_3$} \\
\makebox[0pt][l]{\fboxsep0pt\colorbox{Mygrey} {\strut\hspace*{1.16\linewidth}}}
30. & $\mathcal{E}_1[A_{1,1}^{\oplus 2}A_{3,2}^{\oplus 4}A_{9,1}^{\oplus 2}]$  & $40$  &  \tiny{$\hchi_0 + 2^{19}\hchi_3 + 2^{12}\hchi_3$} &
31. & $\mathcal{E}_1[A_{1,1}A_{5,3}A_{9,1}^{\oplus 2}D_{4,3}]$  & $40$  &  \tiny{$\hchi_0 + 2^{19}\hchi_3 + 2^{12}\hchi_3$} \\
\makebox[0pt][l]{\fboxsep0pt\colorbox{Mywhite} {\strut\hspace*{1.16\linewidth}}}
32. & $\mathcal{E}_1[A_{1,1}A_{7,4}A_{9,1}^{\oplus 2}]$  & $40$  &  \tiny{$\hchi_0 + 2^{19}\hchi_3 + 2^{12}\hchi_3$} & 
33. & $\mathcal{E}_1[A_{9,1}^{\oplus 2}C_{3,2}D_{5,4}]$  & $40$  &  \tiny{$\hchi_0 + 2^{19}\hchi_3 + 2^{12}\hchi_3$} \\
\makebox[0pt][l]{\fboxsep0pt\colorbox{Mygrey} {\strut\hspace*{1.16\linewidth}}}
34. & $\mathcal{E}_1[A_{9,1}^{\oplus 2}D_{6,5}]$  & $40$  &  \tiny{$\hchi_0 + 2^{19}\hchi_3 + 2^{12}\hchi_3$} & 35. & $\mathcal{E}_1[C_{2,1}^{\oplus 3}D_{4,2}^{\oplus 2}E_{7,2}F_{4,1}]$  & $40$  &  \tiny{$\hchi_0 + \hchi_3 + 2^{15}\hchi_3$} \\
\makebox[0pt][l]{\fboxsep0pt\colorbox{Mywhite} {\strut\hspace*{1.16\linewidth}}}
36. & $\mathcal{E}_1[A_{2,1}^{\oplus 2}A_{5,2}^{\oplus 2}E_{7,2}F_{4,1}]$  & $40$  &  \tiny{$\hchi_0 + \hchi_3 + 2^{15}\hchi_3$} & 37. & $\mathcal{E}_1[A_{2,1}E_{6,4}E_{7,2}F_{4,1}]$  & $40$  &  \tiny{$\hchi_0 + \hchi_3 + 2^{15}\hchi_3$} \\
\makebox[0pt][l]{\fboxsep0pt\colorbox{Mygrey} {\strut\hspace*{1.16\linewidth}}}
38. & $\mathcal{E}_1[A_{3,1}^{\oplus 7}A_{7,1}^{\oplus 2}D_{5,1}]$  & $40$  &  \tiny{$\hchi_0 + 2^6\hchi_3 + 2^{17}\hchi_3$} & 39. & $\mathcal{E}_1[A_{3,1}A_{7,1}^{\oplus 2}D_{5,1}D_{5,2}^{\oplus 2}]$  & $40$  &  \tiny{$\hchi_0 + 2^6\hchi_3 + 2^{17}\hchi_3$} \\
\makebox[0pt][l]{\fboxsep0pt\colorbox{Mywhite} {\strut\hspace*{1.16\linewidth}}}
40. & $\mathcal{E}_1[A_{7,1}^{\oplus 2} A_{7,2} C_{3,1}^{\oplus 2} D_{5,1}]$  & $40$  &  \tiny{$\hchi_0 + 2^6\hchi_3 + 2^{17}\hchi_3$} & 41. & $\mathcal{E}_1[A_{7,1}^{\oplus 2} D_{5,1} D_{7,3} G_{2,1}]$  & $40$  &  \tiny{$\hchi_0 + 2^6\hchi_3 + 2^{17}\hchi_3$} \\
\makebox[0pt][l]{\fboxsep0pt\colorbox{Mygrey} {\strut\hspace*{1.16\linewidth}}}
42. & $\mathcal{E}_1[A_{7,1}^{\oplus 2} C_{7,2} D_{5,1}]$  & $40$  &  \tiny{$\hchi_0 + 2^6\hchi_3 + 2^{17}\hchi_3$} & 43. & $\mathcal{E}_1[B_{3,1} B_{4,1} C_{4,1} D_{6,2} D_{8,2}]$  & $40$  &  \tiny{$\hchi_0 + \hchi_3 + 2^{15}\hchi_3$} \\
\makebox[0pt][l]{\fboxsep0pt\colorbox{Mywhite} {\strut\hspace*{1.16\linewidth}}}
44. & $\mathcal{E}_1[A_{4,1}A_{9,2}B_{4,1}D_{8,2}]$  & $40$  &  \tiny{$\hchi_0 + \hchi_3 + 2^{15}\hchi_3$} & 45. & $\mathcal{E}_1[B_{3,1} C_{4,1} C_{6,1}^{\oplus 2} D_{6,2}]$  & $40$  &  \tiny{$\hchi_0 + \hchi_3 + 2^{15}\hchi_3$} \\
\makebox[0pt][l]{\fboxsep0pt\colorbox{Mygrey} {\strut\hspace*{1.16\linewidth}}}
46. & $\mathcal{E}_1[A_{4,1} A_{9,2}  C_{6,1}^{\oplus 2}]$  & $40$  &  \tiny{$\hchi_0 + \hchi_3 + 2^{15}\hchi_3$}  \\
\hline
\hline
\end{longtable}

\section{Conclusions}

The meromorphic theories we have identified in the present work have been summarised in Tables \ref{t4} and \ref{t5}. As mentioned in the Introduction, the proposals in this work are made at a physics level of rigour and considerable evidence provided. It should be possible in future to convert these to a set of mathematically rigorous statements and proofs. We have obtained infinitely many theories that can be thought of as generalisations of 34 non-lattice theories in Ref. \cite{Schellekens:1992db} to arbitrary central charge $c=8N$. The fact that infinitely many generalisations exist is ultimately due to properties of the $B_{r,1}$ and $D_{r,1}$ affine theories, which have three characters for all $r$ and whose modular transformation matrix is periodic in $r$. 

It is interesting to see the Baby Monster module make an appearance in this discussion, in row 1 of Table \ref{t4}. This appears to illustrate a general phenomenon: modules with and without Kac-Moody algebras appear on a similar footing in general meromorphic theories (and presumably in more general RCFTs). This is in contrast to the $c=24$ case where there is one entry (the Monster CFT) that is the extension of modules without a Kac-Moody algebra, namely ${\cal E}_1[{\cal M}(4,3)\,BM]$, while the rest are extensions of only Kac-Moody algebras. Such a perspective may be helpful to derive at least partial classifications for meromorphic theories at $c=32$, even though a complete classification is understood to be virtually impossible due to the enormous number (of order $10^9$) of possible theories.

The present work makes no claim to being complete. The goal has only been to provide several examples of an interesting phenomenon: the prediction of meromorphic theories at higher $c$ starting from coset pairings for meromorphic theories at lower $c$. The present results already suggest many new possibilities, for example we may conjecture that starting at every $c^{\cal H}=8(2p+3)$ (where $p\in \mathbb{N}\cup\{0\}$) there is an infinite series labelled by another parameter $m$ with $(n_1,n_2)=(p+2,p+m+2)$. This series is the extension $\mathcal{E}_1[D_{8p+12,1}D_{8p+8m+12,1}]$. The special case for $p=0$ and arbitrary $m$ corresponds to row 17 of table \ref{t1} which describes a series of meromorphic theories starting at $c^{\cal H}=24$ given by $\mathcal{E}_1[D_{12,1}D_{8m+12,1}]$. Also the special case $p=1$ and $m=0$ can be found in row 1 of table \ref{t3} which describes a  meromorphic theory at $c^{\cal H}=40$ given by, $\mathcal{E}_1[D_{20,1}D_{20,1}]$. Our conjecture subsumes these examples and suggests infinite generalisations thereof. More generally, we have not yet considered cases with multiple $B_{r,1}$ or $D_{r,1}$ factors. We hope to return to all these and many more cases in the future \cite{DGM:upcoming}. 

The recursive nature of the process described here is a potentially very useful feature. If a meromorphic theory is first discovered by any method at any central charge $c=8N$, one can attempt to construct its generalisations at higher values of $c$ using our approach.

Our results provide more evidence for the close relationship between general RCFTs and meromorphic CFT. It is not true, as was implicitly thought earlier (and is still occasionally claimed in the literature), that meromorphic theories constitute some sort of ``exotic'' outliers in the space of all RCFTs. Rather, the most general RCFTs are the ones with extensions of the usual Virasoro and Kac-Moody algebras. The most familiar RCFTs, such as minimal Virasoro and affine theories, are merely unextended special cases of the RCFT landscape which is mostly populated by extensions. Meromorphic theories are such extensions, and are intimately linked to $n$-character RCFT for $n>1$ through coset relations.

\section*{Acknowledgements}

AD and CNG would like to thank Jagannath Santara for collaboration on the papers \cite{Das:2020wsi} and \cite{Das:2021uvd} and for very helpful discussions. AD would like to thank Daniele Dorigoni, Nabil Iqbal and Madalena Lemos for useful discussions on Lie algebras. He would also like to thank Sigma
Samhita for her immense help in the type setting of the tables required for this work. He would also like to express his gratitude to Gabriel Arenas-Henriquez and Jose Cerqueira-sa for helpful discussions on {\it Mathematica}. SM would like to thank Brandon Rayhaun for very helpful discussions, and Keshav Dasgupta and Alex Maloney for hospitality at McGill University, Montreal where part of this work was done. He is also grateful for support from a grant by Precision Wires India Ltd. for String Theory and Quantum Gravity research at IISER Pune. 
\begin{appendix}

\section{Some general features of $B_{r,1}$ and $D_{r,1}$ affine theories}\label{a1}

\begin{itemize}
    \item For $B_{r,1}$ we have the three characters as follows,
    \begin{itemize}
        \item $\chi_0$: Identity (adj. irrep): $[0,\cdots,0]$ (which contains information about the dimension of Lie algebra (= coefficient of $\mathcal{O}(q)$ = $2r^2+r$) and hence is the adjoint irrep of $SO(2r+1)$).
        \item $\chi_{\frac{1}{2}}$: Fundamental irrep: $\sigma_1$ (whose dimension (= coefficient of $\mathcal{O}(q^0)$ term = $2r+1$) matches with that of the fundamental irrep of $SO(2r+1)$). 
        \item $\chi_{\frac{2r+1}{16}}$: Spinor irrep: $\sigma_r$ (whose dimension (= coefficient of $\mathcal{O}(q^0)$ term = $2^r$) matches with that of the spinor irrep of $SO(2r+1)$)\footnote{Dimension of spinorial representation of $SO(n \, \text{odd})=2^{\frac{n-1}{2}}$.} 
    \end{itemize}
    
    \item For $D_{r,1}$ we have the three characters as follows,
    \begin{itemize}
        \item $\chi_0$: Identity (adj. irrep): $[0,\cdots,0]$ (which contains information about the dimension of Lie algebra (= coefficient of $\mathcal{O}(q)$ = $2r^2-r$) and hence is the adjoint irrep of $SO(2r)$)
        \item $\chi_{\frac{1}{2}}$: Fundamental irrep: $\sigma_1$ (whose dimension (= coefficient of $\mathcal{O}(q^0)$ term = $2r$) matches with that of the fundamental irrep of $SO(2r)$) 
        \item $\chi_{\frac{r}{8}}$: Spinor irrep: $\sigma_r \equiv \sigma_{r-1}$ (whose dimension (= coefficient of $\mathcal{O}(q^0)$ term = $2^{r-1}$) matches with that of the spinor irrep of $SO(2r)$)\footnote{Dimension of spinorial representation of $SO(n \, \text{even})=2^{\frac{n}{2}-1}$. In this case (even dimension) we have chiral spinor irreps ($\sigma_r \equiv \sigma_{r-1}$) but at the level of characters they will be treated on equal footing.}
    \end{itemize}    
\end{itemize}

Now note that we have an infinite family whenever we have a factor of $B_{r_1,1}$ or $D_{r_1,1}$ in a theory and the idea is to generate an infinite set by increasing $r_1\to r_1+8k$ (with $k\in\mathbb{Z}$). Now, we can explicitly compute and check the $(d_1,d_2)$ values for upto $c^{\mathcal{H}}=40$ and we observe that we get the same $(d_1,d_2)$ for a given coset pair forming a particular infinite family at every stage of $c^{\cal H}$. This is because the only thing that differs in the bilinear relation of one stage to another for a given infinite family is the rank (in steps of 8) of $B_{r_1,1}$ or $D_{r_1,1}$ factors. However, increasing $r_1\to r_1 + 8$ doesn't change the number of irreps of $B_{r_1,1}$ or $D_{r_1,1}$. Hence, we observe again that the values of $(d_1,d_2)$ at one value of $c^{\cal H}$ will be the same for all other values of $c^{\cal H}$.

The above discussion shows a very crucial point about 3-characters. As we discussed above we were able to generate infinite families of meromorphic theories because each of these theories had a $B_{r_1,1}$ or $D_{r_1,1}$ as their factors and as we know from \cite{Das:2020wsi}, every $B_{r_1,1}$ or $D_{r_1,1}$ has 3 characters (except $D_{4,1}$ which has 2 due to triality). Thus, we could generalise the Schellenkens coset pairs which always resulted in a 9-character theory and from Schellekens we know that the non-$B_{r_1,1}$ or non-$D_{r_1,1}$ factors always admit a unique $3$-character extension.

\section{Relevant admissible character solutions}\label{a3}
In this section we present (for completeness) the relevant admissible character solutions to the $(3, 0)$ MLDE and GHM solutions from \cite{Das:2021uvd} and \cite{Gaberdiel:2016zke} respectively.

\newpage

\setlength\LTleft{0pt}
\setlength\LTright{0pt}
\begin{longtable}{l|cccc|cc||c|cccc|cc}
\caption{Some admissible character solutions to the $(3, 0)$ MLDE and GHM solutions.}
\label{t6}\\
\hline
\hline
\makebox[0pt][l]{\fboxsep0pt\colorbox{Mywhite} {\strut\hspace*{0.96\linewidth}}}
\# & $c$ & $h_1$ & $h_2$ &  $m_1$  & $D_1$ & $D_2$ & \# & $c$ & $h_1$ & $h_2$ &  $m_1$  & $D_1$ & $D_2$  \\
\hline
\hline
\makebox[0pt][l]{\fboxsep0pt\colorbox{Mywhite} {\strut\hspace*{0.96\linewidth}}}
${\bf III_{22}}$ & $\frac{68}{5}$ &  $\frac{4}{5}$ & $\frac{7}{5}$ & $136$ & $119$ & $68$ & ${\bf III_{37}}$ & $\frac{92}{5}$ &  $\frac{6}{5}$ & $\frac{8}{5}$ & $92$ & $1196$ & $299$ \\
\makebox[0pt][l]{\fboxsep0pt\colorbox{Mygrey} {\strut\hspace*{0.96\linewidth}}}
${\bf V_{39}}$ & $20$ &  $\frac{4}{3}$ & $\frac{5}{3}$ & $80$ & $5$ & $4$ & ${\bf III_{45}}$ & $22$ &  $\frac{3}{2}$ & $\frac{7}{4}$ & $66$ & $77$ & $11$ \\
\makebox[0pt][l]{\fboxsep0pt\colorbox{Mywhite} {\strut\hspace*{0.96\linewidth}}}
${\bf III_{50}}$ & $23$ &  $\frac{3}{2}$ & $\frac{15}{8}$ & $23$ & $575$ & $23$ \\
\hline
\hline
\makebox[0pt][l]{\fboxsep0pt\colorbox{Mygrey} {\strut\hspace*{0.96\linewidth}}}
$\text{GHM}_{45}$ & $\frac{45}{2}$ & $\frac{3}{2}$ & $\frac{29}{16}$ & $45$ & $4785$ & $45$ & $\text{GHM}_{86}$ & $\frac{43}{2}$ & $\frac{3}{2}$ & $\frac{27}{16}$ & $86$ & $5031$ & $43$ \\
\makebox[0pt][l]{\fboxsep0pt\colorbox{Mywhite} {\strut\hspace*{0.96\linewidth}}}
$\text{GHM}_{105}$ & $21$ & $\frac{3}{2}$ & $\frac{13}{8}$ & $105$ & $637$ & $21$ & $\text{GHM}_{120}$ & $20$ & $\frac{7}{5}$ & $\frac{8}{5}$ & $120$ & $4$ & $13$ \\
\makebox[0pt][l]{\fboxsep0pt\colorbox{Mygrey} {\strut\hspace*{0.96\linewidth}}}
$\text{GHM}_{123}$ & $\frac{41}{2}$ & $\frac{3}{2}$ & $\frac{25}{16}$ & $123$ & $5125$ & $41$ & $\text{GHM}_{156}$ & $\frac{39}{2}$ & $\frac{3}{2}$ & $\frac{23}{16}$ & $156$ & $5083$ & $39$ \\
\makebox[0pt][l]{\fboxsep0pt\colorbox{Mygrey} {\strut\hspace*{0.96\linewidth}}}
$\text{GHM}_{171}$ & $19$ & $\frac{3}{2}$ & $\frac{11}{8}$ & $171$ & $627$ & $19$ & $\text{GHM}_{185}$ & $\frac{37}{2}$ & $\frac{3}{2}$ & $\frac{21}{16}$ & $185$ & $4921$ & $2368$ \\
\makebox[0pt][l]{\fboxsep0pt\colorbox{Mywhite} {\strut\hspace*{0.96\linewidth}}}
$\text{GHM}_{198}$ & $18$ & $\frac{3}{2}$ & $\frac{5}{4}$ & $198$ & $75$ & $9$ & $\text{GHM}_{210}$ & $\frac{35}{2}$ & $\frac{3}{2}$ & $\frac{19}{16}$ & $210$ & $4655$ & $35$ \\
\makebox[0pt][l]{\fboxsep0pt\colorbox{Mygrey} {\strut\hspace*{0.96\linewidth}}}
$\text{GHM}_{221}$ & $17$ & $\frac{3}{2}$ & $\frac{9}{8}$ & $221$ & $561$ & $17$ & $\text{GHM}_{255}$ & $15$ & $\frac{3}{2}$ & $\frac{7}{8}$ & $255$ & $455$ & $15$ \\
\hline
\hline
\end{longtable}

\end{appendix}

\bibliographystyle{JHEP}

\bibliography{3char}

\end{document}